\documentclass[article,twocolumn,amsmath,amssymb,superscriptaddress,longbibliography]{revtex4-2}
\usepackage[english]{babel}
\usepackage{amssymb}
\usepackage{dcolumn}
\usepackage{bm}
\usepackage{amsmath}
\usepackage{tabularx}
\usepackage{graphicx}        
\usepackage{makecell}                             
\graphicspath{{pict/}{fig2_2D_asymmetric}}

\usepackage{xcolor}
\usepackage{multirow}
\usepackage{dcolumn}
\usepackage{bm}

\usepackage[normalem]{ulem}
\usepackage{comment}
\usepackage{epstopdf}
\usepackage{hhline}

\usepackage[raggedleft]{titlesec}
\titleformat{\section}{\bf\large}{\thesection.\,}{0.24em}{}
\titlespacing{\section}{0cm}{0.5cm}{0cm}

\begin{document}

\onecolumngrid

\title{Topological Dynamics and Correspondences in Composite Exceptional Rings}
\author{Zhoutao Lei}
\affiliation{Guangdong Provincial Key Laboratory of Quantum Metrology and Sensing $\&$ School of Physics and Astronomy, Sun Yat-Sen University (Zhuhai Campus), Zhuhai 519082, China}

\author{Yuangang Deng}
\email{dengyg3@mail.sysu.edu.cn}
\affiliation{Guangdong Provincial Key Laboratory of Quantum Metrology and Sensing $\&$ School of Physics and Astronomy, Sun Yat-Sen University (Zhuhai Campus), Zhuhai 519082, China}
\date{\today}

\begin{abstract}
The study of unconventional phases and elucidation of correspondences between topological invariants and their intriguing properties are pivotal in topological physics. Here, we investigate a complex exceptional ring (CER), composed of a third-order exceptional ring and multiple Weyl exceptional rings, and establish a direct correspondence between Chern numbers and the distinctive behaviors of these structures. We show that band braiding during quasistatic encircling processes correlates with nontrivial Chern numbers, resulting in triple (double) periodic spectra for topologically nontrivial (trivial) middle bands. Moreover, Chern numbers predict mode transfer during dynamical encircling. Experimental schemes for realizing CER in cold atoms are proposed, emphasizing the crucial role of Chern numbers as both measurable quantity and descriptor of exceptional physics in dissipative systems. This discovery broadens topological classifications in non-Hermitian systems, with promising applications in quantum computing and metrology.
\end{abstract}
\maketitle
\clearpage
\
\newpage
\
\newpage

\maketitle

\section*{Introduction}
\noindent
Exceptional points (EPs) are non-Hermitian singularities where both eigenvalues and their corresponding eigenstates coalesce, exhibiting intriguing spectral topology distinct from Hermitian nodal structures which feature only coalescing eigenvalues~\cite{Berry2004,Rotter_2009,Heiss_2012,RevModPhys.90.015001,RevModPhys.93.025002}. These unique spectral degeneracies, fundamental to quantum mechanics, have been observed in diverse platforms such as photonic crystals, acoustic cavities, and solid-state spins~\cite{RevModPhys.93.015005,doi:10.1126/science.aar7709,Lin2023}. EPs catalyze many interesting phenomena, including unconventional
transmission or reflection~\cite{PhysRevLett.103.093902,PhysRevLett.106.213901,Regensburger2012,PhysRevLett.123.214302}, enhanced sensing~\cite{PhysRevLett.112.203901,Chen2017,Hokmabadi2019}, and unusual quantum criticality~\cite{PhysRevX.4.041001,PhysRevLett.119.190401,Li2019,Dora2019,PhysRevLett.123.230401}. It should be emphasized that the quasistatic and dynamical encirclement around EPs exhibits intriguing band braiding~\cite{Wang2021,PhysRevLett.126.010401,PhysRevLett.130.017201,PhysRevLett.130.163001,PhysRevResearch.4.L022064,PhysRevLett.132.243802} and chiral mode transfer~\cite{Doppler2016,Xu2016,Yoon2018,doi:10.1126/science.abl6571,PhysRevLett.126.170506,PhysRevLett.130.157201}, respectively, which signify the unique consequences of nontrivial EPs topology and provide broad applications for novel generation of quantum devices~\cite{RevModPhys.93.015005,doi:10.1126/science.aar7709,Lin2023}. Besides the unique topology of non-Hermitian systems, EPs also inherit some topological properties of spectral degeneracies in Hermitian systems~\cite{RevModPhys.93.015005,RevModPhys.90.015001,RevModPhys.93.025002}.

Recent advancements have extended beyond isolated second-order EPs (2EPs) to discover exotic structures like high-order EPs~\cite{Demange_2012,PhysRevLett.127.186601,PhysRevLett.127.186602,PhysRevLett.130.266901,doi:10.1126/sciadv.adi0732,PhysRevLett.131.100202}, exceptional lines~\cite{PhysRevLett.118.045701,Cerjan2019,PhysRevLett.129.084301,doi:10.1073/pnas.2110018119,Wu2024}, and exceptional nexus~\cite{doi:10.1126/science.abd8872,PhysRevLett.132.253401} leading to swallowtail catastrophe~\cite{Hu2023}, involving more complex coalescing eigenvalues and eigenstates. These exotic exceptional structures enrich the landscape of non-Hermitian physics~\cite{RevModPhys.93.025002,Ding2022,PhysRevB.109.205142} and provide deeper insights into the quantum mechanisms for open systems~\cite{XXYi_2001,Naghiloo2019,Hashemi2022}. Of particular interest, building up a holistic framework governing their non-Hermitian topological phenomena about encirclement process can enhance our understanding of intriguing topological properties and may inspire more applications in non-Hermitian systems~\cite{PhysRevLett.117.107402,Hodaei2017,Zhou2023}. However, {the encircling outcomes obtained around higher-order and multiple EPs are obscure and lacking in characterization. And the} experimental investigations of these higher-order EP geometries remain challenging due to the complexity of constructing  non-Hermitian Hamiltonians with intricate degrees of freedom and symmetry constraints in real quantum systems. 

In this work, we propose an experimental scheme to realize an intriguing composite exceptional ring (CER), which is an exceptional structure composed of one or more distinct exceptional rings. Specifically, the CER consists of a fixed third-order exceptional ring (TER) and several tunable Weyl exceptional rings (WERs), composed of 3EPs and 2EPs, respectively. By leveraging spin-vector-momentum coupling (SVMC) and spin-tensor-momentum coupling (STMC)~\cite{PhysRevLett.120.240401,PhysRevResearch.4.033008,PhysRevLett.129.250501} this CER exhibits a distinct topological charge characterized by quantized Chern numbers. Traditionally associated with quantized Hall conductance~\cite{PhysRevLett.61.2015,RevModPhys.82.1539}, these Chern numbers play a crucial role in dictating the chirality of boundary states, known as the bulk-boundary correspondence~\cite{RevModPhys.93.015005,RevModPhys.90.015001,RevModPhys.93.025002}. Importantly, we establish a distinctive correspondence linking these ``conventional'' topological invariants to the non-Hermitian topology of CER, including quasistatic and dynamical encircling results. We demonstrate that nontrivial bands engage in braiding processes during quasistatic encircling, leading to multiperiod spectra. Concurrently, dynamical encirclement triggers complex chiral-mode-switching behaviors, governed by the Chern numbers of system. This intricate behavior not only exemplifies the tunable topological charges of the CER but also extends our understanding of topological phenomena beyond conventional two-band systems. {Especially, this correspondence can serve as a guiding principle for analyzing encirclement outcomes, which are invariably obscure and challenging to encapsulate in multiband systems, as well as for designing exceptional structures to attain targeted encirclement results.} Furthermore, we show that CER and its encircling dynamics can be experimentally realized with cold atoms. Our work opens new avenues for exploring non-Hermitian topological properties of other complex exceptional structures and for extracting topological invariants in dissipative systems with crystal symmetry~\cite{PhysRevX.7.041069,Bradlyn2017,Po2017}.

\begin{figure*}[!htp]
\includegraphics[width=1.9\columnwidth]{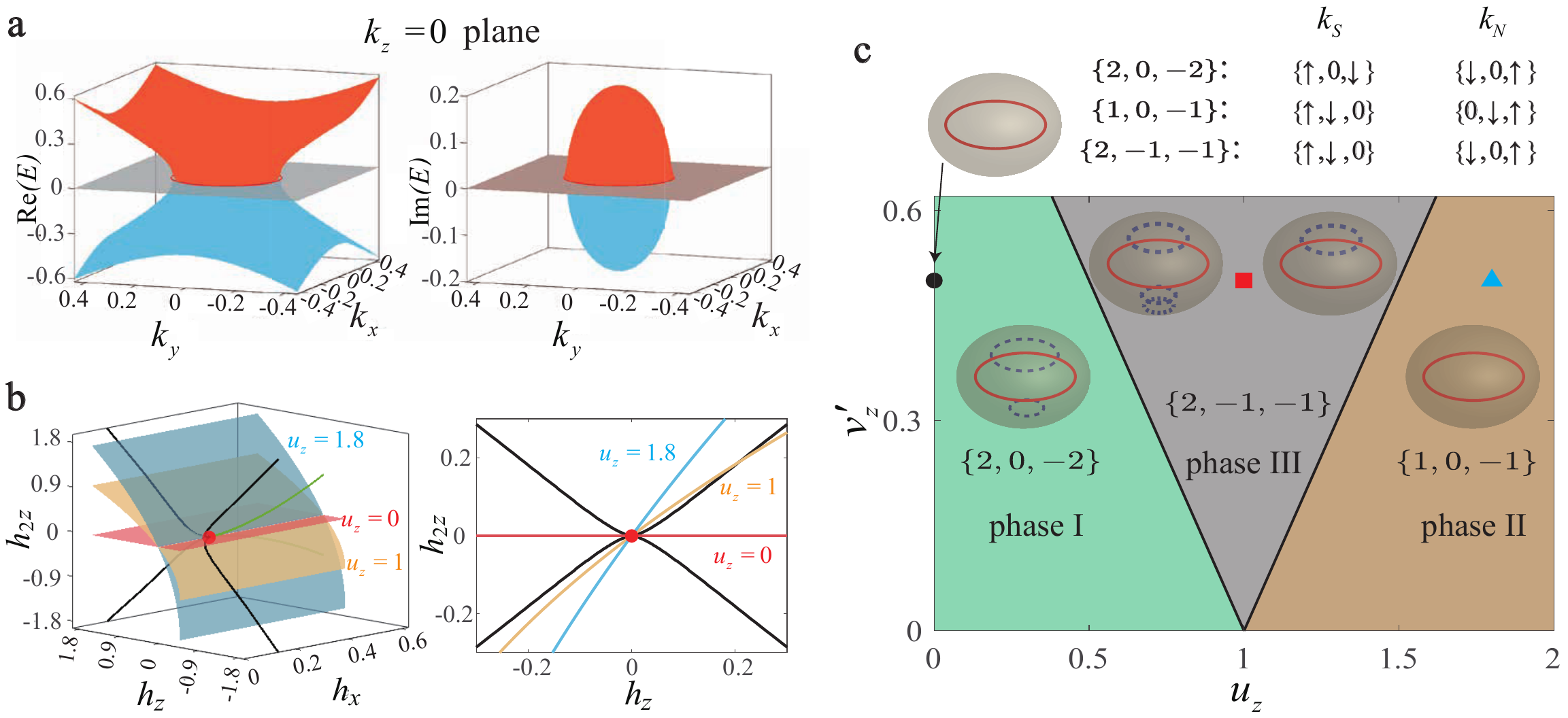}
\caption{\label{fig1} \textbf{Evolution of CER and corresponding Chern number.} \textbf{a} The typical spectrum in $k_z=0$ plane. \textbf{b} The exceptional nexus in $h_{x}$-$h_{z}$-$h_{2z}$ space (left panel) with the surfaces defined by varying $\mathbf{k}$ and its projection (right panel). The parameters used in defining surfaces are indexed by circles, triangles, and squares in \textbf{c}. The structure comprises exceptional lines formed by 2EPs, denoted by green and black lines, which converge at a 3EP marked by a red circle. It is emphasized that only the black lines, which do not lie in $h_z=0$ plane, can intersect with the surfaces. \textbf{c} The phase diagram with Chern numbers $\{\mathcal{C}_{-1},\mathcal{C}_{0},\mathcal{C}_{1}\}$ and evolution of CER upon inclusion of STMC and SVMC. 
The pseudospin configurations of each band at the
south (north) poles of the integral surface, namely the brown spherical surface, are
shown below ${k}_S$ (${k}_N$).
{The typical distribution of exceptional rings has been illustrated for each phase, where red solid (blue dashed) circle indexes the TER (WERs).} In all panels, $\gamma=0.2$ is maintained.}
\end{figure*}

\section*{Results}
\noindent
\textbf{Hamiltonian and Symmetries.}
Building upon recent exploration of triply degenerate points {representing a \sout{new} type of quasiparticle excitations without high-energy counterparts and} characterized by distinct topological charges~\cite{PhysRevLett.120.240401,PhysRevResearch.4.033008,PhysRevLett.129.250501} (also detailed in the Supplementary Note 1), we begin with the following non-Hermitian Hamiltonian incorporating system dissipation
\begin{eqnarray}\label{CER} 
H(\mathbf{k})=\sum_{\alpha=x,y,z}k_{\alpha}\hat{F}_{\alpha}+u_zk_z\hat{F}_{z}^2+v'_zk^2_z\hat{F}_{z}+i\gamma\hat{F}_{z}, \label{Ham}
\end{eqnarray}
where $\hat{F}_{\alpha}$ represents the spin-1 matrix within a pseudospin-1 framework and $k_{\alpha}$ denotes the effective momentum. The terms $k_{\alpha}\hat{F}_{\alpha}$ and $u_zk_z\hat{F}_{z}^2$ manifest as SVMC and STMC~\cite{PhysRevLett.120.240401,PhysRevResearch.4.033008,PhysRevLett.129.250501} respectively. Besides the linear dispersion, $v'_zk^2_z\hat{F}_{z}$ corresponds to the quadratic dispersion component of SVMC, which an be emerged in ultracold pseudospin-1 atomic gases confined within optical lattices~\cite{PhysRevResearch.4.033008}. The term $i\gamma\hat{F}_{z}$ introduces a non-Hermitian interaction associated with asymmetric dissipation acting on pseudospin states spin-$\uparrow$ and spin-$\downarrow$.

The Hamiltonian (\ref{Ham}) possesses several intriguing symmetries, that significantly influence the generation of exceptional structures. Firstly, it possesses rotational symmetry around the $z$-axis, defined by $\hat{\mathcal{C}}_{z}H(\mathbf{k})\hat{\mathcal{C}}^{-1}_{z}=H[\mathcal{R}(\mathbf{e}_z,{\phi})\mathbf{k}]$ with $\hat{\mathcal{C}}_{z}=e^{-i\phi\hat{F}_z}$ and $\mathcal{R}(\mathbf{e}_z,{\phi})$ being element of ${\rm SO}(3)$ group. This symmetry imposes a spatial arrangement of EPs into rings. Additionally, the Hamiltonian preserves both {parity}-time ($\mathcal{PT}$) symmetry~\cite{PhysRevLett.80.5243} and pseudochirality symmetry~\cite{Wu2024} in $k_z=0$ plane. This is articulated by the relations $e^{i\pi\hat{F}_{x}}H(k_x,k_y,0)e^{-i\pi\hat{F}_{x}}=H^{*}(k_x,k_y,0)$ and $e^{i\pi\hat{F}_{z}}H(k_x,k_y,0)e^{-i\pi\hat{F}_{z}}=-H^{\dag}(k_x,k_y,0)$. In light of these two symmetries, a significant TER emerges in this plane, characterized by the equation $k_x^2+k_y^2=\gamma^2$, delineating a phase transition boundary with respect to $\mathcal{PT}$ symmetry breaking, as depicted in Fig.~\ref{fig1}a. Beyond this plane, the presence of these two symmetries will vanish associated with the absence of 3EP, although WER composed of 2EPs may appear. Furthermore, in the absence of quadratic SVMC interaction ($v'_z=0$), the Hamiltonian also exhibits $\mathcal{CM}_{x}$ symmetry, expressed as $e^{i\pi\hat{F}_{z}}H(k_x,k_y,k_z)e^{-i\pi\hat{F}_{z}}=-H^{*}(k_x,-k_y,-k_z)$. This symmetry represents a hybridization of particle-hole and mirror symmetries. Under $\mathcal{CM}_{x}$ symmetry, WERs necessarily appear in symmetric pairs distributed about the $k_z=0$ plane.
\\

\noindent\textbf{Topological composite exceptional ring.}
To delineate the formation and spatial positioning of WERs, we analyze the Hamiltonian in Eq.~\eqref{CER} within the $k_y=0$ plane, leveraging the rotational $\hat{\mathcal{C}}_{z}$ symmetry. In this configuration, pairs of symmetric EPs extends to form an exceptional ring in 3D space. Specifically, within this setting, the Hamiltonian simplifies to
\begin{eqnarray}\label{CERky0}
H_1(\mathbf{k})=h_x\hat{F}_{x}+h_z\hat{F}_{z}+h_{2z}\hat{F}^2_{z}+i\gamma\hat{F}_{z},
\end{eqnarray}
where $h_{x}=k_x$, $h_{z}=k_z+v'_zk^2_z$, and $h_{2z}=u_zk_z$. The spectrum analysis of $H_1(\mathbf{k})$ reveals six two-fold exceptional lines in the $h_{x}$-$h_{z}$-$h_{2z}$ parameter space, converging at a 3EP [marked by red circle at coordinates $(|\gamma|,0,0)$], as illustrated in Fig.~\ref{fig1}b. This unique configuration is known as an exceptional nexus~\cite{doi:10.1126/science.abd8872}. Projecting this nexus onto the $h_{z}$-$h_{2z}$ plane, results in distinctive mappings of the exceptional lines: two (green lines) project onto the line $h_{z}=0$, while the others (black lines) map onto curves approaching the asymptotes $h_{2z}=\pm h_{z}$ as $|h_{2z}|$ tends to infinity. 
 
The parameters $h_{x}$, $h_{z}$, and $h_{2z}$ within $H_1(\mathbf{k})$ depend on $k_x$ and $k_z$, defining a surface in the parameter space. This surface invariably intersects the 3EP at $k_x=\pm\gamma$ and $k_z=0$, correlating with the emergence of TER described in Eq.~\eqref{CER}. Specifically, under the purely linear dispersion from SVMC interaction ($u'_z=v_z=0$), the parameter surface remains flat with  $h_{2z}=0$, intersecting the exceptional nexus solely at the 3EP. After incorporating STMC interaction ($u_z\neq0$ and $v'_z=0$), the initially flat surface inclines with a slope of $u_z$, allowing intersections with the exceptional lines (black lines) at two symmetric points. These intersections herald the emergence of an additional pair of WERs {composed of 2EPs.} with opposite momenta $k_z$, consistent with the above discussed forementioned symmetries. As strength of STMC increases, these WERs move away from the 3EP towards higher values of $|k_z|$ within the range $(0,1)$, and eventually vanish when $u_z>1$, due to the asymptotic relationship $h_{2z}=\pm h_{z}$ inherent to these exceptional lines.

When both quadratic SVMC and STMC are incorporated into the Hamiltonian, the parameter surface curves, projecting as a parabola in the $h_{z}$-$h_{2z}$ plane, corresponding to the location of WERs becoming asymmetric about $k_z$. The number of WERs within $|k_z|\in(0,1)$ remains symmetric about $k_z=0$ plane for $0<u_z<1-v'_z$, becoming asymmetric as $u_z$ falls within the range $1-v'_z<u_z<1+v'_z$. Outside these ranges, the WERs completely disappear when $u_z>1+v'_z$ or $u_z=0$. Figure~\ref{fig1}b illustrates the outcomes for various values of $u_z$ while fixing $v'_z$. Clearly, intersections are confined to the 3EP (3EP and one 2EP) for $u_z>1+v'_z$ and $u_z=0$ ($1-v'_z<u_z<1+v'_z$) scenarios. The evolution of CER are summarized in Fig.~\ref{fig1}c, with detailed discussions available in Supplementary Note 2.

Further analysis reveals that the CER possesses a nontrivial topological charge characterized by Chern number
\begin{eqnarray}\label{Chern}
\mathcal{C}_n\equiv\frac{1}{2\pi}\oint_{\mathcal{S}}\mathbf{\Omega}_n\cdot d\mathbf{S},
\end{eqnarray}
where $\mathbf{\Omega}_n(\mathbf{k})=\mathbf{\nabla}_{\mathbf{k}}\times
\langle\psi_n(\mathbf{k})|i\mathbf{\nabla}_{\mathbf{k}}|\psi_n(\mathbf{k})\rangle$ denotes the Berry curvature of the $n$th band. Here the integral surface ${\mathcal{S}}$ encloses CER. 

\begin{figure}[!htp]
\includegraphics[width=0.9\columnwidth]{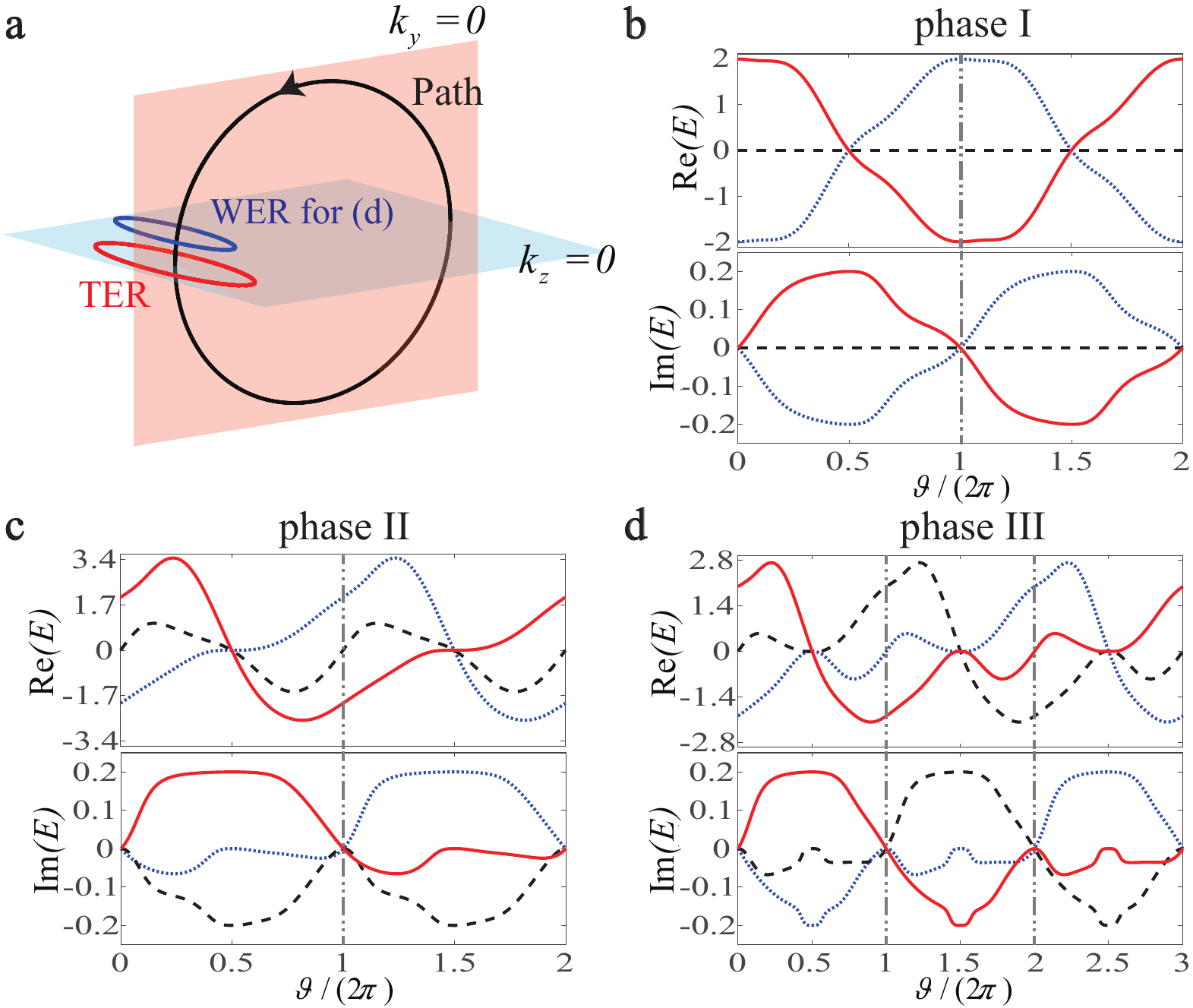}
\caption{\label{fig2} \textbf{Spectral interchange in the quasistatic encircling process around CERs.} \textbf{a} The typical closed path (black circles) defined in Eq.~\eqref{path} with radius $R=1$.  \textbf{b}-\textbf{d} The spectra responses during the quasistatic encircling of CER under different parameter settings. {Here, we use vertical gray dashed lines to index points completing encirclements.} The parameters are $v'_{z}=0.5$ and $\gamma=0.2$, while $u_{z}=0$ for \textbf{b}, $1.8$ for \textbf{c}, and $1$ for \textbf{d}. In these cases, the CER includes only one TER (red circle in \textbf{a}) in \textbf{b} and \textbf{c}, and an additional WER (blue circle in \textbf{a}) in \textbf{d}.}
\end{figure}

In Fig.~\ref{fig1}c, the CER is classified into three distinct topological phases according to its Chern number. Specifically, in the regime dominated by linear SVMC ($u_z<1-v'_z$) and indexed as phase {\rm I}, the Chern number for CER is given by $\mathcal{C}_n=-2n$, regardless of whether the CER supports paired WERs ($u_z\neq0$) or not ($u_z=0$). Conversely, in the phase {\rm II} where STMC dominated $u_z>1+v'_z$, the CER encompasses only one TER and exhibits a Chern number of $\mathcal{C}_n=-n$.
In these scenarios, only two bands exhibit nontrivial Chern numbers, akin to the characteristics of a Weyl point and WER. By contrast in the intermediated regime, $1-v'_z<u_z<1+v'_z$ the phase {\rm III} is generated by the interplay between STMC and quadratic SVMC, yielding Chern number configuration of $\{2,-1,-1\}$. This phase showcases all three nontrivial bands, highlighting complex topological dynamics induced by the combination of STMC and SVMC.

Under the rotational symmetry $\hat{\mathcal{C}}_{z}$, eigenstates along the rotation axis $k_x=k_y=0$ also serve as eigenstates of $\hat{\mathcal{C}}_{z}$, thus preserving their bare spin state. Consequently, the topological charge of the CER can be extracted from the eigenstates at the poles of ${\mathcal{S}}$, as depicted at the top of Fig.~\ref{fig1}c. This leads to a simplified expression for the Chern number~\cite{PhysRevResearch.4.033008,PhysRevLett.129.250501} (see Method section):
\begin{eqnarray}\label{spin}
\mathcal{C}_n=\langle\psi_n(\mathbf{k}_S)|\hat{F}_z|\psi_n(\mathbf{k}_S)\rangle-\langle\psi_n(\mathbf{k}_N)|\hat{F}_z|\psi_n(\mathbf{k}_N)\rangle,
\end{eqnarray}
where $\mathbf{k}_S$ ($\mathbf{k}_N$) represents the south (north) pole of the integral surface $\mathcal{S}$. Specifically, the $E_0$-band does not occur spin-flipping for phases {\rm I} and {\rm II}, while it undergoes a transition from pseudospin-$\downarrow$ to pseudospin-$0$ for phases {\rm III} with three nontrivial bands. This distinctive relationship between spin features and topological charges of the CER leads to diverse topological phenomena during the encircling process, even when the encircling path remains fixed, as discussed subsequently.
\\

\noindent\textbf{Quasistatic encircling process.}
A defining characteristic of non-Hermitian topology at EPs is the exchange of eigenstates and associated eigenvalues along a quasistatic path encircling the EP, resulting in a nontrivial braiding structure~\cite{Wang2021,PhysRevLett.126.010401,PhysRevLett.130.017201,PhysRevLett.130.163001,PhysRevResearch.4.L022064,Bouhon2020,PhysRevB.102.115135,PhysRevLett.125.053601,PhysRevLett.132.243802} .
To explore this property, we define a closed path encircling the CER described by 
\begin{eqnarray}\label{path}
\mathbf{k}=R[(1+\cos\vartheta)\cos\phi,(1+\cos\vartheta)\sin\phi,\sin\vartheta],
\end{eqnarray}
where $\vartheta$ varies from $0$ to $2\pi$ with $\phi$ and $R$ being constants. It's worth noting that the choice of $\phi$ does not affect the spectral behavior due to the rotational symmetry $\hat{\mathcal{C}}_{z}$. For simplicity, we set $\phi=0$, confining the path to the $k_y=0$ plane, as shown in Fig.~\ref{fig2}a. Crucially, the integer period point of this closed path $\mathbf{k}_{0}$ falls within the region of unbroken $\mathcal{PT}$ symmetry. This aspect is intricately linked to the topological charge of the CER. In scenarios with a trivial middle band, the branches corresponding to the highest and lowest energies at the initial point [$E_{\pm1}(\mathbf{k}_{0})$] intersect, while the third branch returns to its initial state, as illustrated in Figs.~\ref{fig2}b and \ref{fig2}c. In contrast, all branches will swap with each others when CER hosts three nontrivial bands, giving rise to the triple period beyond the two-band non-Hermitian system, as shown in Fig.~\ref{fig2}d. Thereby a nontrivial correspondence is established:
\begin{eqnarray}\label{braidper}
N({\mathcal{C}_n\neq0})=T_{\vartheta}/2\pi,
\end{eqnarray}
where $N({\mathcal{C}_n\neq0})$ and $T_{\vartheta}/2\pi$ denotes the number of nontrivial bands and the multiple of the spectra period. This relationship inherently reflects a distinctive spin texture associated with different topological charges of the CER. During this quasistatic process, each branch evolves into specific bare spin states at the half period point around $\mathbf{k}_N$. Subsequently, during the transition from momentum near $\mathbf{k}_N$ to momentum near $\mathbf{k}_S$, the associated eigenstate predominantly maintains this bare spin state considering $H\approx i\gamma\hat{F}_z$.  As the path returns to the $\mathcal{PT}$ symmetric point, these bare spin states evolve into eigenstates with real energy. Consequently, band with nontrivial Chern number associated with different bare spin states $\mathbf{k}_N$ and $\mathbf{k}_S$, switch after one period, thereby verifying the correspondence in Eq.~\eqref{braidper}. 
{We also calculate the Berry phases of the branch over a completed spectra period~\cite{PhysRevLett.118.045701}.  It is found that the two braiding branches shown in Fig.~\ref{fig2}c possess nontrivial Berry phases of $\pm\pi$ whereas the leftmost one has a trivial Berry phase. Additionally, all branches depicted in Figs.~\ref{fig2}b and \ref{fig2}d exhibit zero Berry phases.}
Additional examples and the confirmation of this quasistatic topological property,determined by the Chern number of the CER, are provided in the Supplementary Note 3.
\\

\begin{figure}[!htp]
\includegraphics[width=1\columnwidth]{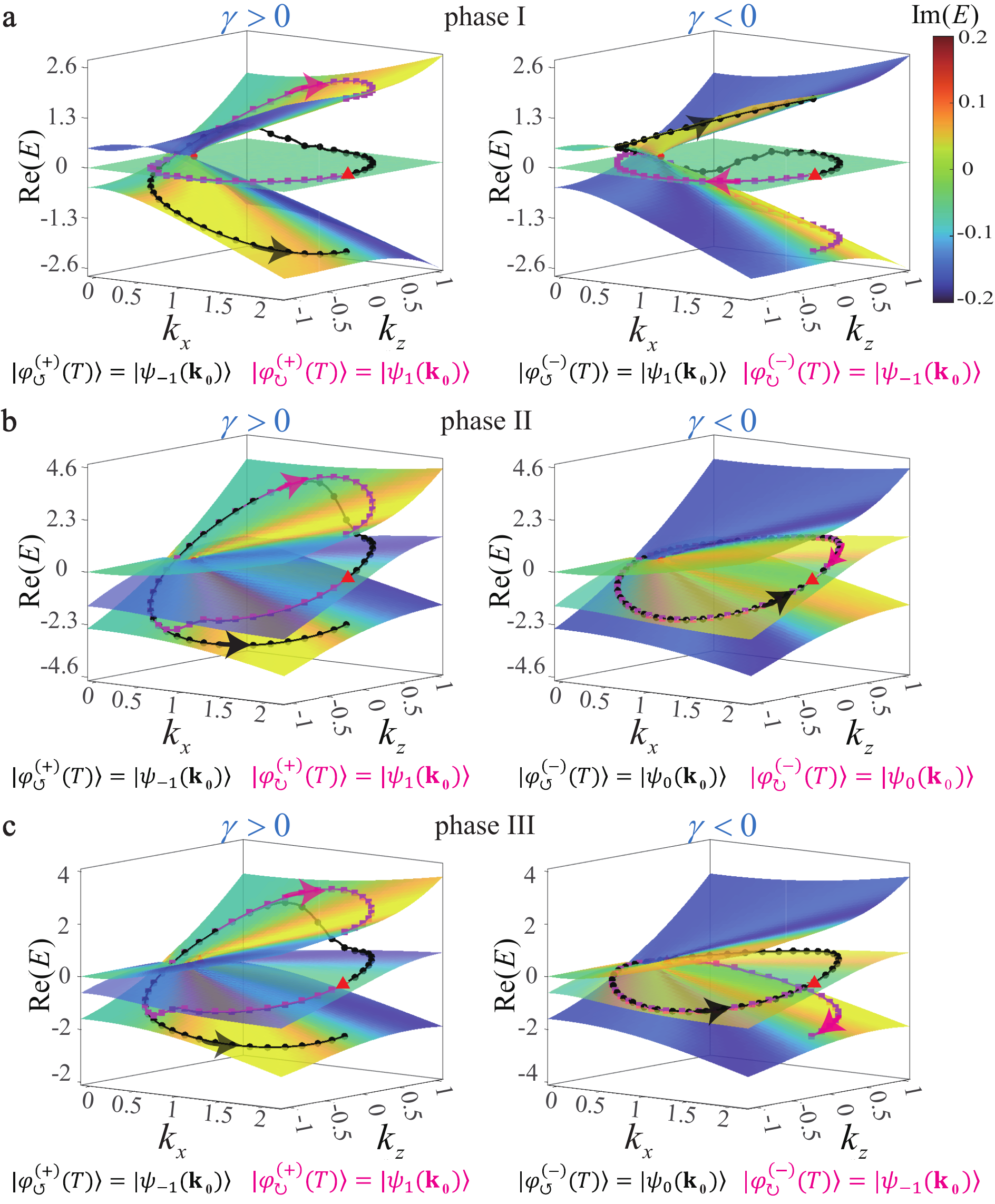}
\caption{\label{fig3} \textbf{Dynamical encirclement of CERs with varying directions and dissipations.} The expectation energy during {counterclockwise} ({clockwise}) encirclement process is depicted by black lines with circles (magenta lines with squares). The energy of initial state is marked with a red triangle. Parameters are set to $v'_{z}=0.5$, $|\gamma|=0.2$, and $|\omega|=0.02\pi$ for all panels. The specific value of $u_{z}$ is $u_{z}=0$ for \textbf{a}, $1.8$ for \textbf{b}, and $1$ for \textbf{c}, matching those in Fig.~\ref{fig1}{b}.}
\end{figure}

\noindent\textbf{Dynamical encircling process.}
The diverse topological characteristics of CER lead to more intriguing phenomena during dynamic encirclement. Specifically, we dynamically tune the parameter $\vartheta=\omega t$ in Eq.~\eqref{path} and the evolution of the state is governed by:
\begin{eqnarray}\label{dyna}
|\varphi^{(\pm)}_{\circlearrowleft/\circlearrowright}(t)\rangle=e^{-i\int_0^tH[\mathbf{k}(t)]dt}|\varphi(0)\rangle,
\end{eqnarray}
where the subscript denotes clockwise and counterclockwise dynamics associated with $\omega<0$ and $\omega>0$, respectively, while the superscript indexes the sign of dissipation ${\gamma}$. Importantly, the Hamiltonian exhibits $\mathcal{PT}$ symmetry after periods $\mathbf{k}(mT)=\mathbf{k}_0$, where $T=2\pi/|\omega|$ and $m$ is an integer. Particularly, the Hamiltonian approximate reduces to $H[\mathbf{k}(mT)]\approx2R\hat{F}_x$ when dissipation is significantly weaker compared to the coupling strength at the initial time $|\gamma/2R|\ll1$.

Figure~\ref{fig3} displays the evolution of the state on energy surface during one encircling process, where the outcome is independent of the initial state configuration. As depicted, the encircling process consistently results in chiral behavior that remains invariant with respect to the topological charge of the CER when $\gamma>0$, producing a pair of biorthogonal states. Specifically, under {counterclockwise} ({clockwise}) dynamic, the state transitions to the branch with the lowest (highest) real energy: $|\varphi^{(+)}_{\circlearrowleft}(T)\rangle=|\psi_{-1}(\mathbf{k}_{0})\rangle$ [$|\varphi^{(+)}_{\circlearrowright}(T)\rangle=|\psi_{1}(\mathbf{k}_{0})\rangle$]. Conversely, CERs with different topological charges exhibit distinct encircling dynamics when $\gamma<0$. In phase {\rm I}, the results for {clockwise} and {counterclockwise} dynamics are reversed, as shown in the right panel of Fig.~\ref{fig3}a, analogous to the behavior observed in non-Hermitian two-band systems. While the altering of dissipation will produce symmetric outcome for phase {\rm II}: $|\varphi^{(-)}_{\circlearrowleft}(T)\rangle=|\varphi^{(-)}_{\circlearrowright}(T)\rangle=|\psi_{0}(\mathbf{k}_{0})\rangle$, shown in Fig.~\ref{fig3}b. Despite both scenarios involving one TER, their distinct topological charges result in different encircling dynamical behaviors. Moreover, in phase {\rm III}, where the CER hosts all nontrivial bands, unique chiral dynamics are evidently illustrated by $|\varphi^{(-)}_{\circlearrowleft}(T)\rangle=|\psi_{0}(\mathbf{k}_{0})\rangle$ and $|\varphi^{(-)}_{\circlearrowright}(T)\rangle=|\psi_{-1}(\mathbf{k}_{0})\rangle$, contrasting with other scenarios, as shown in Fig.~\ref{fig3}c.

\begin{figure}[!htp]
\includegraphics[width=1\columnwidth]{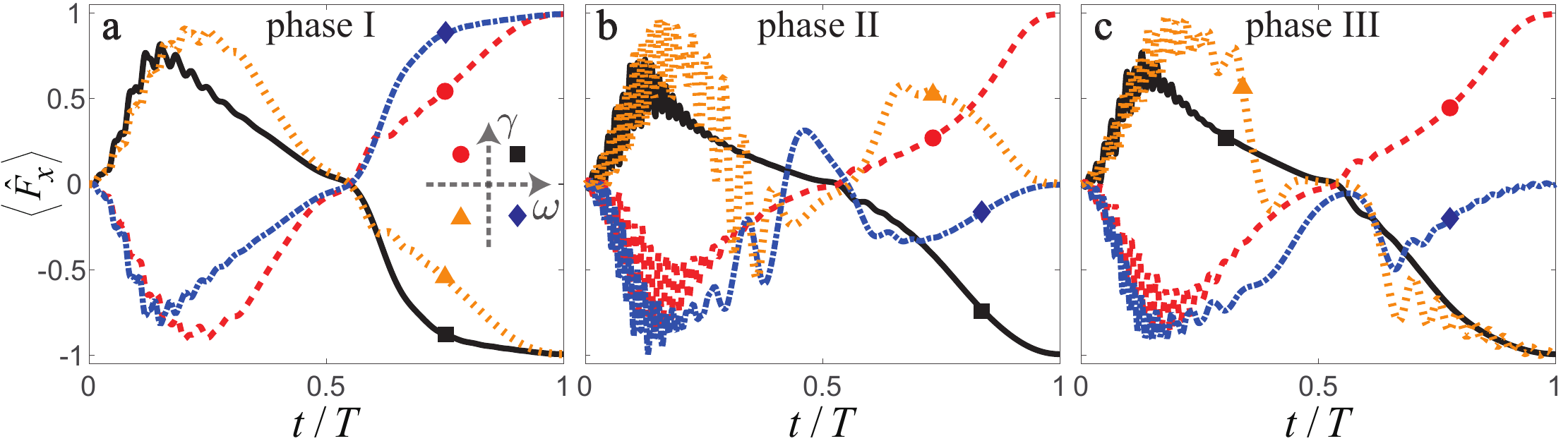}
\caption{\label{Newfig4}\textbf{Observable during dynamic encircling of CER in cold atoms.} The observable $\langle\hat{F}_x\rangle$ defined in Eq.~\eqref{obser} throughout the dynamic encirclement of CERs with distinct topological charges. The initial state is a Guassian wavepacket centered at zero momentum with a width of $\sigma=4$. Parameters are fixed at $v'_{z}=0.5$, $|\gamma|=0.2$, $|\omega|=0.02\pi$ and $R=1$ for all panels. Specific values of $u_{z}$ are $0$ for \textbf{a}, $1.8$ for \textbf{b}, and $1$ for \textbf{c}. Results for different signs of $\gamma$ and $\omega$ are distinguished by different line and symbol colors (red circle, black square, orange triangle, and blue diamond), as indicated in the inset of \textbf{a}.}
\end{figure}

The dynamical results of encircling processes around CER are governed by the Chern number, as demonstrated with additional examples in the Supplementary Note 3 and explained by Eq.~\eqref{spin}. Notably, the evolution of the system is  prominently influenced by the dissipation term around the midpoint of the cycle ($t=T/2$), causing the states to converge towards the bare spin state $|\uparrow\rangle$ ($|\downarrow\rangle$) for $\gamma>0$ ($\gamma<0$). Consequently, the momentum evolves from near $\mathbf{k}_{S}$ ($\mathbf{k}_{N}$) to $\mathbf{k}_{0}$ in the {counterclockwise} ({clockwise}) encirclement process, corresponding to the state evolving adiabatically. Thus, the results of encircling are determined by which branch aligns with $|\uparrow\rangle$ ($|\downarrow\rangle$) at $\mathbf{k}_{S}$ and $\mathbf{k}_{N}$ in the {counterclockwise} and {clockwise} encirclement processes under positive (negative) dissipation, respectively. Considering the relationship given in Eq.~\eqref{spin}, the correspondence between dynamic encircling process and Chern number is further established by the equation 
\begin{eqnarray}\label{dynacorr}
\mathcal{C}_n=\sum_{\alpha={\pm}}\sum_{w=\circlearrowright/\circlearrowleft}\alpha r_w|\langle\psi_{n}(\mathbf{k}_{0})|\varphi^{(\alpha)}_{w}(T)\rangle|^2,
\end{eqnarray}
with $r_{\circlearrowright}=-r_{\circlearrowleft}=1$.
This pivotal relationship provides a systematic approach to analyzing state transitions in different exceptional structures and extracting topological invariants from dissipative topological phases. {Actually, this dynamical correspondence is established via the bridge that connects topological invariants and symmetry indicators at higher symmetric momenta. Thus, this approach can be applied to topological phases defined by other topological invariants as well.
Given that a large number of topological crystalline phases have been discovered through symmetry-based indicators~\cite{PhysRevX.7.041069,Bradlyn2017,Po2017}, it should be noted that some of these phases cannot be characterized by conventional topological invariants and even lack the bulk-boundary correspondence.
Our results can be extended to these systems and established dynamical correspondence will be conducive to extracting their topological features, which can be achieved by introducing suitable dissipation and investigating the dynamical process.}\\

\noindent\textbf{Experimental scheme with cold atomic gases.}
The observed CER with its dynamical encirclement can be realized for ultracold gases trapped in 2D square optical lattice. The experimental setup involves three hyperfine states in the spin-1 ground manifold, manipulated by Raman-assisted spin-orbit couplings. Additionally, the dissipation for $|\uparrow\rangle$ and $|\downarrow\rangle$ states can be generated by using a resonant optical beam~\cite{Li2019}. To proceed further, the lattice Hamiltonian in momentum space under periodic boundary conditions is given by (further details can be found in Method section):
\begin{eqnarray}\label{Mo_Hamil}
{H}_0(\mathbf{k})=\sum_{\alpha=x,y,z}d_{\alpha}(\mathbf{k})\hat{F}_{\alpha}+d'_z(\mathbf{k})\hat{F}^2_{z}+\epsilon(\mathbf{k})\hat{I},
\end{eqnarray}
where $d_{x}(\mathbf{k})=\sin k_x$, $d_{y}(\mathbf{k})=\sin k_y$, $d_{z}(\mathbf{k})=\delta+i\gamma$, $d'_{z}(\mathbf{k})=\delta_2-4t_u\cos k_x-4t_u\cos k_y$, and $\epsilon(\mathbf{k})=2t_u\cos k_x+2t_u\cos k_y$ with $t_u$ being the strength of spin-independent hopping. And $\delta$ and $\delta_2$ are two-photon detunings dependent on a tunable parameter $k_z$: $\delta=k_z+v_z'k^2_z$ and $\delta_2=u_zk_z+8t_u$. The effective Hamiltonian near $\mathbf{k}=0$ corresponds to the formulation presented in Eq.~\eqref{CER}. The square geometry is taken for this lattice system, where non-Hermitian skin effect can be prevented~\cite{Zhang2022,PhysRevLett.131.036402} due to the four-fold rotation symmetry $e^{-i\pi\hat{F}_z/2}H_0(k_x,k_y)e^{i\pi\hat{F}_z/2}=H_0(-k_y,k_x)$. Real and imaginary components of the spectrum can be observed through frequency and decay rate of Rabi oscillations in cold atom system~\cite{Li2019}, thereby enabling the detection of CER.

To facilitate the dynamical encircling process, the incorporation of on-site spin-flipping dynamics is crucial, which can be described by the Hamiltonian:
\begin{eqnarray}\label{Mo_Hamil_dy}
{H}_{\rm dy}(\mathbf{k},t)={H}_0(\mathbf{k})+t_x\hat{F}_x,
\end{eqnarray}
where $t_x$ denotes the strength of on-site spin-flipping, adjusted to control the central momentum $k_x$ of CER. For constructing the dynamical encircling process, we modulate $t_x=R[1+\cos(\omega t)]$ and adjust $k_z=R\sin(\omega t)$. Here, we assume a deep lattice where the nearest-neighbor tunneling is negligible ($t_u=0$). The initial state is prepared as a zero-momentum Gaussian wavepacket in real space with a width $\sigma$, and in spin space as the state $|0\rangle$. To monitor the dynamical process, an observable related to the spin component of zero momentum state $|v(t)\rangle=\langle \mathbf{k}=0|\varphi(t)\rangle$ is introduced:
\begin{eqnarray}\label{obser}
\langle\hat{F}_x\rangle=\frac{\langle v(t)|\hat{F}_x|v(t)\rangle}{\langle v(t)|v(t)\rangle}.
\end{eqnarray}
This observable can be directly measured using spin-resolved time-of-flight images in cold atom experiments~\cite{doi:10.1126/science.aaf6689}. After completing one encircling period, $\langle\hat{F}_x\rangle$ is expected to approximate values close to $-1$, $0$, or $1$, as shown in Fig.~\ref{Newfig4}, reflecting the characteristics of the continuous Hamiltonian dynamics depicted in Fig.~\ref{fig3}, considering $H(\mathbf{k}_0)\approx2R\hat{F}_x$. These configurations not only highlight the versatility of the experimental setup but also deepen the understanding of dynamical topological phenomena in non-Hermitian systems.

\section*{Conclusion}
\noindent
We have explored an intriguing exceptional structure named as CER, consisting of a TER and a variable ensemble of WERs {based on recent research into exotic band degeneracy points associated with unconventional quasiparticle excitations}. By manipulating the interplay between STMC and quadratic SVMC, this CER exhibits a range of topological charges, characterized by Chern number. Importantly, we have established the clear link between these Chern numbers and the topological properties embed during the encirclement process. The emergence of the CER, driven by STMC and quadratic SVMC, enables intricate band braiding and state transitions that transcend conventional two-band systems. This relationship underscores the profound link between {conventional} topological invariants and dynamic quantum phenomena {unique to non-Hermitian systems}, enhancing our comprehension of topological physics in non-Hermitian systems. {Specifically, one can utilize the topological charge to analyze and even design the encirclement physics of other exceptional structures. Conversely, it is also possible to measure the topological charge through encirclement outcomes. The interplay between this dynamical correspondence and other characteristics of topological phases might give rise to more fascinating phenomena, which we will explore in the future.} Furthermore, we have proposed achievable experimental setups utilizing cold atoms in optical lattices to realize the CER. The insights gained from our study may pave the way for further exploration of unique exceptional structures and hold promise for diverse applications in designing quantum devices based on non-Hermitian topologies.

\section*{Methods}
\noindent
\textbf{Chern numbers and spin textures}\label{appA}
In this section we discuss the relationship between spin textures and the Chern number of CER. We choose the integration surface $\mathcal{S}$ in the Eq.~\eqref{Chern} as a spherical surface enclosing CER, centered at the origin with polar (azimuth) angle $\theta$ ($\phi$). Due to the rotation symmetry, Hamiltonian $H$ from Eq.~\eqref{CER} satisfies $H(\theta,\phi)=e^{-i\phi\hat{F}_z}H(\theta,0)e^{i\phi\hat{F}_z}$, and then corresponding eigenstates of $n$th band can be expressed as:
\begin{eqnarray}\label{states} |\psi_n(\theta,\phi)\rangle&=&e^{-i\phi\hat{F}_z}|\psi_n(\theta,0)\rangle,\nonumber\\
|\psi_n(\theta,0)\rangle&=&(\psi_{n,\uparrow},\psi_{n,0},\psi_{n,\downarrow})^T,
\end{eqnarray}
where $|\psi_n(\theta,0)\rangle$ is a smooth and single valued function of the polar angle $\theta$. The Berry connection is then given by:
\begin{eqnarray}\label{conne} A_{\theta}=\langle\psi_n(\theta,\phi)|i\partial_{\theta}\psi_n(\theta,\phi)\rangle=i\sum_{\alpha=\uparrow,0,\downarrow}\psi^*_{n,\alpha}\partial_{\theta}\psi_{n,\alpha},\nonumber\\
A_{\phi}=\langle\psi_n(\theta,\phi)|i\partial_{\phi}\psi_n(\theta,\phi)\rangle=\langle\psi_n(\theta,0)|\hat{F}_z|\psi_n(\theta,0)\rangle.\nonumber\\
\end{eqnarray}
Notifying $|\psi_n(\theta,0)\rangle$ is independent of $\phi$, the Berry curvature will be:
\begin{eqnarray}\label{curva}
\Omega_{\theta\phi}=\partial_{\theta}A_{\phi}-\partial_{\phi}A_{\theta}=\partial_{\theta}[\langle\psi_n(\theta,0)|\hat{F}_z|\psi_n(\theta,0)\rangle].
\end{eqnarray}
As a result, the Chern number of CER is expressed as:
\begin{eqnarray}\label{Cherncurva}
\mathcal{C}_n=\frac{1}{2\pi}\int_0^{2\pi}d\phi\int_0^{\pi}d\theta\Omega_{\theta\phi}=[\langle\psi_n(\theta,0)|\hat{F}_z|\psi_n(\theta,0)\rangle]|^{\theta=\pi}_{\theta=0},\nonumber\\
\end{eqnarray}
which corresponds precisely to Eq.\eqref{spin}.

It is noteworthy that Eq.~\eqref{Cherncurva} holds true for both Hermitian~\cite{PhysRevResearch.4.033008,PhysRevLett.129.250501} and non-Hermitian system. Previous studies have demonstrate that the Chern number of {triply degenerate point} can be revealed by measuring spin texture along the rotation axis $k_x=k_y=0$, where it equals the spin-flipping behavior at the origin (the location of {triply degenerate point}). In the Hermitian system, the band index can be reliably determined based on their local energies. In contrast, the band index is determined by the $\mathcal{PT}$ symmetric momenta $\mathbf{k}_{0}$ in present non-Hermitian system.
\\
\noindent\textbf{Derivation of the lattice Hamiltonian for cold atoms.}
In this section, we derive the lattice Hamiltonian for cold atoms in detail.

\begin{figure}[!htp]
\includegraphics[width=0.8\columnwidth]{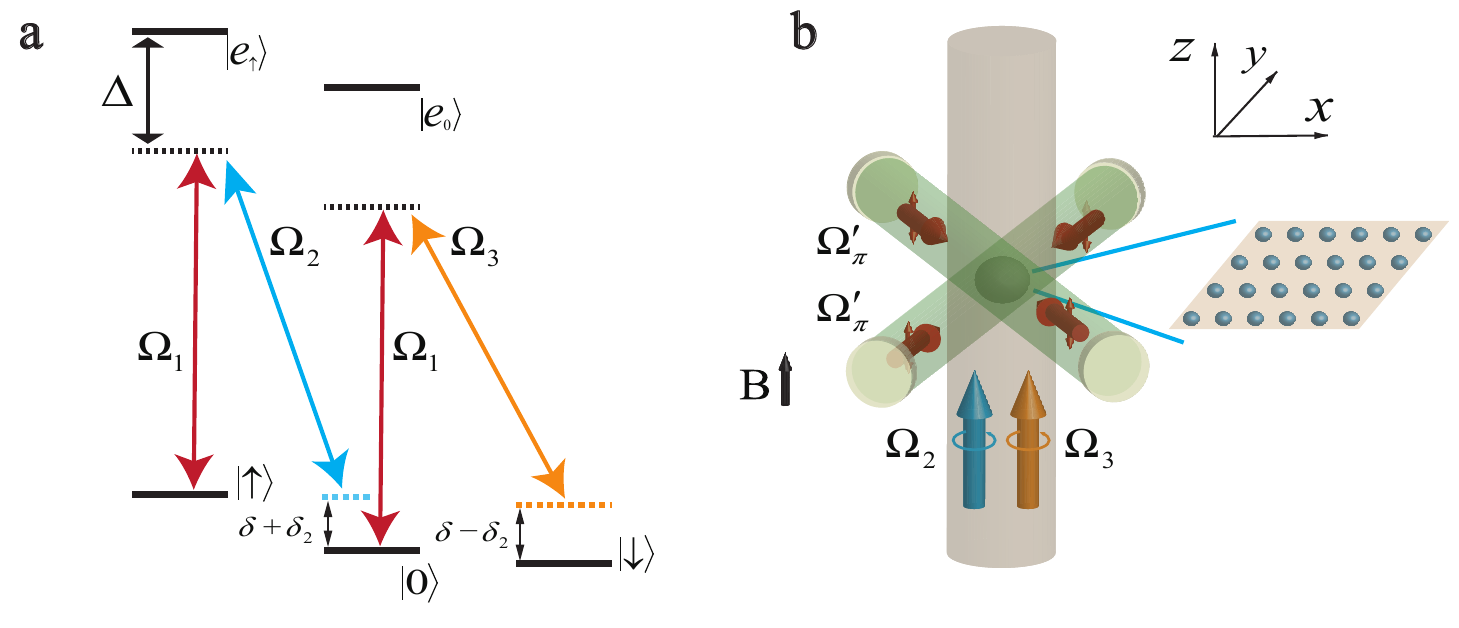}
\caption{\label{subfig_atom_laser}\textbf{Scheme for realizing CER in cold atoms.} \textbf{a} Involved five sublevels. \textbf{b} Illustration of laser configurations and optical lattices. The atoms are trapped in a 2D optical lattice. The red lasers represent 
$\pi$-polarized standing-wave lasers that couple the state $|\sigma\rangle$ to $|e_{\sigma}\rangle$ with $\sigma=\uparrow,0$, while the blue (orange) laser represents a $\sigma$-polarized standing-wave laser along $z$ direction that couples $|0\rangle\leftrightarrow|e_{\uparrow}\rangle$ ($|e_0\rangle\leftrightarrow|\downarrow\rangle$).}
\end{figure}

As discussed in above and shown in Fig.~\ref{subfig_atom_laser}, we consider an ultracold gas subjected to a bias magnetic along $z$ axis and trapped in a square optical lattice $\mathcal{U}_{\rm{ol}}(\mathbf{r})=-U_{ol}[\cos^2(k_Lx)+\cos^2(k_Ly)]$. We focus on five relevant magnetic Zeeman sublevels, including three ground states and two electronic excited states, as shown in Fig.~\ref{subfig_atom_laser}a. The atomic transition frequency between the ground and excited states is denoted by $\omega_a$. Under the influence of the magnetic field, the linear and quadratic Zeeman shifts for the ground (excited) states are $\omega_Z$ and $\omega_q$ ($\omega'_Z$ and $\omega'_q$), respectively.

To manipulate the atoms, we introduce two $\pi$-polarized standing-wave lasers, represented by the red lines in Fig.~\ref{subfig_atom_laser}, which couple the state $|\sigma\rangle$ to $|e_{\sigma}\rangle$ with $\sigma=\uparrow,0$. The detuning between the atomic and laser frequencies is $\Delta=\omega_a-\omega_L$  with $\omega_L$ being the laser frequency. The Rabi frequencies of two standing-wave lasers are $\Omega'_{\pi}\sin(k_Lx-k_Ly)$ and $i\Omega'_{\pi}\sin(k_Lx+k_Ly)$, respectively, giving rise to the total Rabi frequency $\Omega_{1}({\bf r})=\Omega_{\pi}[\sin(k_Lx)\cos(k_Ly)+i\cos(k_Lx)\sin(k_Ly)]$ with $\Omega_{\pi}=\sqrt{2}e^{i\pi/4}\Omega'_{\pi}$. Additionally, a $\sigma$-polarized standing-wave laser along $z$ direction with Rabi frequency $\Omega_{\sigma}$ couples $|0\rangle\leftrightarrow|e_{\uparrow}\rangle$ ($|e_0\rangle\leftrightarrow|\downarrow\rangle$) with frequency $\omega_L+\Delta\omega_L$ ($\omega_L-\Delta\omega_L'$), as indicated by the blue (orange) line in Fig.~\ref{subfig_atom_laser}. Moreover, additional radio frequency pulses generate effective dissipation for $|\uparrow\rangle$ and $|\downarrow\rangle$ states.

When the atom-light detuning $\Delta$ is large, we can adiabatically eliminate the excited states to derive the effective Hamiltonian within the ground state manifold~\cite{PhysRevResearch.4.033008}:
\begin{eqnarray}
{\boldsymbol{h}}_0=&&\frac{{\mathbf{p}}^{2}}{2M}+M_x\mathbf{(r)}\hat{F}_{x}+M_y\mathbf{(r)}\hat{F}_{y}\nonumber\\
&&+(\delta+i\gamma)\hat{F}_{z}+\delta_2\hat{F}^2_z+\mathcal{U}_{\rm{ol}}(\mathbf{r})\hat{I},
\end{eqnarray}
where $M_x(r)=\Omega\sin(k_Lx)\cos(k_Ly)$, and $M_y(r)=\Omega\cos(k_Lx)\sin(k_Ly)$ with $\Omega=-\sqrt{2}\Omega_{\pi}\Omega_{\sigma}/\Delta$. Here $\delta=\omega_z+\frac{\Delta\omega'_L-\Delta\omega_L}{2}+\frac{|\omega_{\sigma}|^2}{2\Delta}$ and $\delta_2=-\omega_q-\frac{\Delta\omega'_L+\Delta\omega_L}{2}-\frac{3|\omega_{\sigma}|^2}{2\Delta}$ act as tunable effective Zeeman fields, and $\gamma$ represents the strength of dissipation.

Incorporating nearest-neighbor hopping between lowest $s$ orbits, we obtain the lattice Hamiltonian
\begin{eqnarray}\label{Eq.realspacelat}
{H_0}=\sum_{\mathbf{j},\alpha=x,y}&&[it_0\hat{\psi}^{\dag}_{\mathbf{j}}\hat{F}_{\alpha}\hat{\psi}_{\mathbf{j} +\mathbf{1}_\alpha} +t_u\hat{\psi}^{\dag}_{\mathbf{j}}(2\hat{F}^2_{z}-\hat{I})\hat{\psi}_{\mathbf{j}+\mathbf{1}_{\alpha}} + {\rm H.c.}]\nonumber\\
&&+\sum_{\mathbf{j}}[(\delta+\delta_2+i\gamma)\hat{n}_{\mathbf{j},\uparrow}-(\delta-\delta_2+i\gamma)\hat{n}_{\mathbf{j},\downarrow}],\nonumber\\
\end{eqnarray}
where $\hat{\psi}_{\mathbf{j}}=[\hat{a}_{\mathbf{j},\uparrow},\hat{a}_{\mathbf{j},0},\hat{a}_{\mathbf{j},\downarrow}]^T$
 with $\hat{a}_{\mathbf{j},\sigma}$ is the annihilation operator for the $\sigma$ component at site $\mathbf{j}$, and $\hat{n}_{\mathbf{j},\sigma}=\hat{a}^{\dag}_{\mathbf{j},\sigma}\hat{a}_{\mathbf{j},\sigma}$. $t_u$ is the spin-independent hopping and $t_0$ is the strength of spin-flipping hopping.

To further process the system, a gauge transformation, $\hat{a}_{\mathbf{j},0}\rightarrow (-1)^{(m+n)}i\hat{a}_{\mathbf{j},0}$, has been implemented~\cite{PhysRevResearch.4.033008}. Upon Fourier transformation, the Hamiltonian in momentum space takes the form as (the lattice constant is taken as unit $a\equiv1/k_L\equiv1$)
\begin{eqnarray}\label{Mo_HamilT0}
{H}_0(\mathbf{k})=\sum_{\alpha=x,y,z}d_{\alpha}(\mathbf{k})\hat{F}_{\alpha}+d'_z(\mathbf{k})\hat{F}^2_{z}+\epsilon(\mathbf{k})\hat{I},
\end{eqnarray}
where the coefficients are defined as follows:  $d_{x}(\mathbf{k})=2t_0\sin k_x$, $d_{y}(\mathbf{k})=2t_0\sin k_y$, $d_{z}(\mathbf{k})=\delta+i\gamma$, $d'_{z}(\mathbf{k})=\delta_2-4t_u\cos k_x-4t_u\cos k_y$, and $\epsilon(\mathbf{k})=2t_u\cos k_x+2t_u\cos k_y$. Additionally, the parameters $\delta$ and $\delta_2$ act analogously to effective Zeeman fields. Setting $\delta=2t_0k_z+4v_z't^2_0k^2_z$ and $\delta_2=2u_zt_0k_z+8t_u$, which vary with the parameter $k_z$, the effective Hamiltonian expanded around $k_x=k_y=0$ is given by:
\begin{eqnarray}\label{effe_Mo_Hamil}
{H}_{\rm CER}=2t_0\sum_{\alpha=x,y,z}k_{\alpha}\hat{F}_{\alpha}+(4v_z't^2_0k^2_z+i\gamma)\hat{F}_{z}+2u_zt_0k_z\hat{F}^2_{z}.\nonumber\\
\end{eqnarray}
Upon setting $2t_0$ as the energy unit ($2t_0\equiv1$), ${H}_{\rm CER}$ corresponds to the Hamiltonian of CER in Eq.\eqref{CER}.

As discussed in above, on-site spin-flipping is introduced to realize the dynamic encircling process around CER.
This effect is achieved by introducing two $\pi$-polarized standing-wave lasers that couple $|\sigma\rangle$ to $|e_{\sigma}\rangle$ with $\sigma=\uparrow,0$ at the frequency $\omega_L$, as the red one in Fig.~\ref{subfig_atom_laser}. The Rabi frequencies of these lasers are $\Omega''_{\pi}\cos(k_Lx-k_Ly)$ and $\Omega''_{\pi}\cos(k_Lx+k_Ly)$, yielding the total Rabi frequency of $2\Omega''_{\pi}\cos(k_Lx)\cos(k_Ly)$. As a result, the momentum space Hamiltonian is modified to
\begin{eqnarray}\label{Mo_Hamil_dyR}
{H}_{\rm dy}={H}_0+t_x\hat{F}_x,
\end{eqnarray}
whose momentum representation is detailed in Eq.\eqref{Mo_Hamil_dy}. By adjusting $t_x$, $\delta$, and $\delta_2$, the dynamical encircling process can be realized.

We assume the lattice depth is sufficiently deep to suppress tunneling ($t_u=0$). Other parameters are tuned as
\begin{eqnarray}\label{paracold}
t_x&=&2t_0R[1+\cos(\omega t)],\nonumber\\\delta&=&2t_0R\sin(\omega t)+4v'_zt^2_0R^2\sin^2(\omega t),\nonumber\\\delta_2&=&2u_zt_0R\sin(\omega t)
\end{eqnarray}
during the dynamical process. The initial state is prepared as a Gaussian wavepacket:
\begin{eqnarray}\label{Gua}
|\varphi(0)\rangle=\frac{1}{\sqrt{\mathcal{N}}}\sum_{m,n}e^{-[(m-m_0)^2+(n-n_0)]^2/4\sigma^2}|m,n\rangle\otimes|0\rangle,\nonumber\\
\end{eqnarray}
where $\mathcal{N}$ is the normalization factor, $m$ ($n$) is the lattice index along $x$ ($y$) direction with $m_0$ ($n_0$) denoting the center position of wavepacket, and $\sigma$ is the width of Gaussian wavepacket. This setup allows for the simulation of the dynamical encircling process. Future investigations will explore additional topological features, such as boundary states within this lattice model. 

\section*{Data availability}
\noindent
Raw numerical data from the plots presented are available from the authors upon request.

\section*{Code availability}
\noindent
Though not essential to the central conclusions of this work, computer codes for generating our figures are available from the authors upon reasonable request.

\section*{References}

\begin{thebibliography}{72}%
\makeatletter
\providecommand \@ifxundefined [1]{%
 \@ifx{#1\undefined}
}%
\providecommand \@ifnum [1]{%
 \ifnum #1\expandafter \@firstoftwo
 \else \expandafter \@secondoftwo
 \fi
}%
\providecommand \@ifx [1]{%
 \ifx #1\expandafter \@firstoftwo
 \else \expandafter \@secondoftwo
 \fi
}%
\providecommand \natexlab [1]{#1}%
\providecommand \enquote  [1]{``#1''}%
\providecommand \bibnamefont  [1]{#1}%
\providecommand \bibfnamefont [1]{#1}%
\providecommand \citenamefont [1]{#1}%
\providecommand \href@noop [0]{\@secondoftwo}%
\providecommand \href [0]{\begingroup \@sanitize@url \@href}%
\providecommand \@href[1]{\@@startlink{#1}\@@href}%
\providecommand \@@href[1]{\endgroup#1\@@endlink}%
\providecommand \@sanitize@url [0]{\catcode `\\12\catcode `\$12\catcode
  `\&12\catcode `\#12\catcode `\^12\catcode `\_12\catcode `\%12\relax}%
\providecommand \@@startlink[1]{}%
\providecommand \@@endlink[0]{}%
\providecommand \url  [0]{\begingroup\@sanitize@url \@url }%
\providecommand \@url [1]{\endgroup\@href {#1}{\urlprefix }}%
\providecommand \urlprefix  [0]{URL }%
\providecommand \Eprint [0]{\href }%
\providecommand \doibase [0]{https://doi.org/}%
\providecommand \selectlanguage [0]{\@gobble}%
\providecommand \bibinfo  [0]{\@secondoftwo}%
\providecommand \bibfield  [0]{\@secondoftwo}%
\providecommand \translation [1]{[#1]}%
\providecommand \BibitemOpen [0]{}%
\providecommand \bibitemStop [0]{}%
\providecommand \bibitemNoStop [0]{.\EOS\space}%
\providecommand \EOS [0]{\spacefactor3000\relax}%
\providecommand \BibitemShut  [1]{\csname bibitem#1\endcsname}%
\let\auto@bib@innerbib\@empty
\bibitem [{\citenamefont {Berry}(2004)}]{Berry2004}%
  \BibitemOpen
  \bibfield  {author} {\bibinfo {author} {\bibfnamefont {M.~V.}\ \bibnamefont
  {Berry}},\ }\bibfield  {title} {\bibinfo {title} {Physics of nonhermitian
  degeneracies},\ }\href {https://doi.org/10.1023/B:CJOP.0000044002.05657.04}
  {\bibfield  {journal} {\bibinfo  {journal} {Czech. J. Phys.}\ }\textbf
  {\bibinfo {volume} {54}},\ \bibinfo {pages} {1039} (\bibinfo {year}
  {2004})}\BibitemShut {NoStop}%
\bibitem [{\citenamefont {Rotter}(2009)}]{Rotter_2009}%
  \BibitemOpen
  \bibfield  {author} {\bibinfo {author} {\bibfnamefont {I.}~\bibnamefont
  {Rotter}},\ }\bibfield  {title} {\bibinfo {title} {A non-hermitian hamilton
  operator and the physics of open quantum systems},\ }\href
  {https://doi.org/10.1088/1751-8113/42/15/153001} {\bibfield  {journal}
  {\bibinfo  {journal} {J. Phys. A}\ }\textbf {\bibinfo {volume} {42}},\
  \bibinfo {pages} {153001} (\bibinfo {year} {2009})}\BibitemShut {NoStop}%
\bibitem [{\citenamefont {Heiss}(2012)}]{Heiss_2012}%
  \BibitemOpen
  \bibfield  {author} {\bibinfo {author} {\bibfnamefont {W.~D.}\ \bibnamefont
  {Heiss}},\ }\bibfield  {title} {\bibinfo {title} {The physics of exceptional
  points},\ }\href {https://doi.org/10.1088/1751-8113/45/44/444016} {\bibfield
  {journal} {\bibinfo  {journal} {J. Phys. A}\ }\textbf {\bibinfo {volume}
  {45}},\ \bibinfo {pages} {444016} (\bibinfo {year} {2012})}\BibitemShut
  {NoStop}%
\bibitem [{\citenamefont {Armitage}\ \emph {et~al.}(2018)\citenamefont
  {Armitage}, \citenamefont {Mele},\ and\ \citenamefont
  {Vishwanath}}]{RevModPhys.90.015001}%
  \BibitemOpen
  \bibfield  {author} {\bibinfo {author} {\bibfnamefont {N.~P.}\ \bibnamefont
  {Armitage}}, \bibinfo {author} {\bibfnamefont {E.~J.}\ \bibnamefont {Mele}},\
  and\ \bibinfo {author} {\bibfnamefont {A.}~\bibnamefont {Vishwanath}},\
  }\bibfield  {title} {\bibinfo {title} {Weyl and dirac semimetals in
  three-dimensional solids},\ }\href
  {https://doi.org/10.1103/RevModPhys.90.015001} {\bibfield  {journal}
  {\bibinfo  {journal} {Rev. Mod. Phys.}\ }\textbf {\bibinfo {volume} {90}},\
  \bibinfo {pages} {015001} (\bibinfo {year} {2018})}\BibitemShut {NoStop}%
\bibitem [{\citenamefont {Lv}\ \emph {et~al.}(2021)\citenamefont {Lv},
  \citenamefont {Qian},\ and\ \citenamefont {Ding}}]{RevModPhys.93.025002}%
  \BibitemOpen
  \bibfield  {author} {\bibinfo {author} {\bibfnamefont {B.~Q.}\ \bibnamefont
  {Lv}}, \bibinfo {author} {\bibfnamefont {T.}~\bibnamefont {Qian}},\ and\
  \bibinfo {author} {\bibfnamefont {H.}~\bibnamefont {Ding}},\ }\bibfield
  {title} {\bibinfo {title} {Experimental perspective on three-dimensional
  topological semimetals},\ }\href
  {https://doi.org/10.1103/RevModPhys.93.025002} {\bibfield  {journal}
  {\bibinfo  {journal} {Rev. Mod. Phys.}\ }\textbf {\bibinfo {volume} {93}},\
  \bibinfo {pages} {025002} (\bibinfo {year} {2021})}\BibitemShut {NoStop}%
\bibitem [{\citenamefont {Bergholtz}\ \emph {et~al.}(2021)\citenamefont
  {Bergholtz}, \citenamefont {Budich},\ and\ \citenamefont
  {Kunst}}]{RevModPhys.93.015005}%
  \BibitemOpen
  \bibfield  {author} {\bibinfo {author} {\bibfnamefont {E.~J.}\ \bibnamefont
  {Bergholtz}}, \bibinfo {author} {\bibfnamefont {J.~C.}\ \bibnamefont
  {Budich}},\ and\ \bibinfo {author} {\bibfnamefont {F.~K.}\ \bibnamefont
  {Kunst}},\ }\bibfield  {title} {\bibinfo {title} {Exceptional topology of
  non-hermitian systems},\ }\href
  {https://doi.org/10.1103/RevModPhys.93.015005} {\bibfield  {journal}
  {\bibinfo  {journal} {Rev. Mod. Phys.}\ }\textbf {\bibinfo {volume} {93}},\
  \bibinfo {pages} {015005} (\bibinfo {year} {2021})}\BibitemShut {NoStop}%
\bibitem [{\citenamefont {Miri}\ and\ \citenamefont
  {Al\`{u}}(2019)}]{doi:10.1126/science.aar7709}%
  \BibitemOpen
  \bibfield  {author} {\bibinfo {author} {\bibfnamefont {M.-A.}\ \bibnamefont
  {Miri}}\ and\ \bibinfo {author} {\bibfnamefont {A.}~\bibnamefont {Al\`{u}}},\
  }\bibfield  {title} {\bibinfo {title} {Exceptional points in optics and
  photonics},\ }\href {https://doi.org/10.1126/science.aar7709} {\bibfield
  {journal} {\bibinfo  {journal} {Science}\ }\textbf {\bibinfo {volume}
  {363}},\ \bibinfo {pages} {eaar7709} (\bibinfo {year} {2019})}\BibitemShut
  {NoStop}%
\bibitem [{\citenamefont {Lin}\ \emph {et~al.}(2023)\citenamefont {Lin},
  \citenamefont {Tai}, \citenamefont {Li},\ and\ \citenamefont
  {Lee}}]{Lin2023}%
  \BibitemOpen
  \bibfield  {author} {\bibinfo {author} {\bibfnamefont {R.}~\bibnamefont
  {Lin}}, \bibinfo {author} {\bibfnamefont {T.}~\bibnamefont {Tai}}, \bibinfo
  {author} {\bibfnamefont {L.}~\bibnamefont {Li}},\ and\ \bibinfo {author}
  {\bibfnamefont {C.~H.}\ \bibnamefont {Lee}},\ }\bibfield  {title} {\bibinfo
  {title} {Topological non-hermitian skin effect},\ }\href
  {https://doi.org/10.1007/s11467-023-1309-z} {\bibfield  {journal} {\bibinfo
  {journal} {Frontiers of Physics}\ }\textbf {\bibinfo {volume} {18}},\
  \bibinfo {pages} {53605} (\bibinfo {year} {2023})}\BibitemShut {NoStop}%
\bibitem [{\citenamefont {Guo}\ \emph {et~al.}(2009)\citenamefont {Guo},
  \citenamefont {Salamo}, \citenamefont {Duchesne}, \citenamefont {Morandotti},
  \citenamefont {Volatier-Ravat}, \citenamefont {Aimez}, \citenamefont
  {Siviloglou},\ and\ \citenamefont
  {Christodoulides}}]{PhysRevLett.103.093902}%
  \BibitemOpen
  \bibfield  {author} {\bibinfo {author} {\bibfnamefont {A.}~\bibnamefont
  {Guo}}, \bibinfo {author} {\bibfnamefont {G.~J.}\ \bibnamefont {Salamo}},
  \bibinfo {author} {\bibfnamefont {D.}~\bibnamefont {Duchesne}}, \bibinfo
  {author} {\bibfnamefont {R.}~\bibnamefont {Morandotti}}, \bibinfo {author}
  {\bibfnamefont {M.}~\bibnamefont {Volatier-Ravat}}, \bibinfo {author}
  {\bibfnamefont {V.}~\bibnamefont {Aimez}}, \bibinfo {author} {\bibfnamefont
  {G.~A.}\ \bibnamefont {Siviloglou}},\ and\ \bibinfo {author} {\bibfnamefont
  {D.~N.}\ \bibnamefont {Christodoulides}},\ }\bibfield  {title} {\bibinfo
  {title} {Observation of $\mathcal{P}\mathcal{T}$-symmetry breaking in complex
  optical potentials},\ }\href {https://doi.org/10.1103/PhysRevLett.103.093902}
  {\bibfield  {journal} {\bibinfo  {journal} {Phys. Rev. Lett.}\ }\textbf
  {\bibinfo {volume} {103}},\ \bibinfo {pages} {093902} (\bibinfo {year}
  {2009})}\BibitemShut {NoStop}%
\bibitem [{\citenamefont {Lin}\ \emph {et~al.}(2011)\citenamefont {Lin},
  \citenamefont {Ramezani}, \citenamefont {Eichelkraut}, \citenamefont
  {Kottos}, \citenamefont {Cao},\ and\ \citenamefont
  {Christodoulides}}]{PhysRevLett.106.213901}%
  \BibitemOpen
  \bibfield  {author} {\bibinfo {author} {\bibfnamefont {Z.}~\bibnamefont
  {Lin}}, \bibinfo {author} {\bibfnamefont {H.}~\bibnamefont {Ramezani}},
  \bibinfo {author} {\bibfnamefont {T.}~\bibnamefont {Eichelkraut}}, \bibinfo
  {author} {\bibfnamefont {T.}~\bibnamefont {Kottos}}, \bibinfo {author}
  {\bibfnamefont {H.}~\bibnamefont {Cao}},\ and\ \bibinfo {author}
  {\bibfnamefont {D.~N.}\ \bibnamefont {Christodoulides}},\ }\bibfield  {title}
  {\bibinfo {title} {Unidirectional invisibility induced by
  $\mathcal{P}\mathcal{T}$-symmetric periodic structures},\ }\href
  {https://doi.org/10.1103/PhysRevLett.106.213901} {\bibfield  {journal}
  {\bibinfo  {journal} {Phys. Rev. Lett.}\ }\textbf {\bibinfo {volume} {106}},\
  \bibinfo {pages} {213901} (\bibinfo {year} {2011})}\BibitemShut {NoStop}%
\bibitem [{\citenamefont {Regensburger}\ \emph {et~al.}(2012)\citenamefont
  {Regensburger}, \citenamefont {Bersch}, \citenamefont {Miri}, \citenamefont
  {Onishchukov}, \citenamefont {Christodoulides},\ and\ \citenamefont
  {Peschel}}]{Regensburger2012}%
  \BibitemOpen
  \bibfield  {author} {\bibinfo {author} {\bibfnamefont {A.}~\bibnamefont
  {Regensburger}}, \bibinfo {author} {\bibfnamefont {C.}~\bibnamefont
  {Bersch}}, \bibinfo {author} {\bibfnamefont {M.-A.}\ \bibnamefont {Miri}},
  \bibinfo {author} {\bibfnamefont {G.}~\bibnamefont {Onishchukov}}, \bibinfo
  {author} {\bibfnamefont {D.~N.}\ \bibnamefont {Christodoulides}},\ and\
  \bibinfo {author} {\bibfnamefont {U.}~\bibnamefont {Peschel}},\ }\bibfield
  {title} {\bibinfo {title} {Parity-time synthetic photonic lattices},\ }\href
  {https://doi.org/10.1038/nature11298} {\bibfield  {journal} {\bibinfo
  {journal} {Nature (London)}\ }\textbf {\bibinfo {volume} {488}},\ \bibinfo
  {pages} {167} (\bibinfo {year} {2012})}\BibitemShut {NoStop}%
\bibitem [{\citenamefont {Wang}\ \emph {et~al.}(2019)\citenamefont {Wang},
  \citenamefont {Fang}, \citenamefont {Mao}, \citenamefont {Jing},\ and\
  \citenamefont {Li}}]{PhysRevLett.123.214302}%
  \BibitemOpen
  \bibfield  {author} {\bibinfo {author} {\bibfnamefont {X.}~\bibnamefont
  {Wang}}, \bibinfo {author} {\bibfnamefont {X.}~\bibnamefont {Fang}}, \bibinfo
  {author} {\bibfnamefont {D.}~\bibnamefont {Mao}}, \bibinfo {author}
  {\bibfnamefont {Y.}~\bibnamefont {Jing}},\ and\ \bibinfo {author}
  {\bibfnamefont {Y.}~\bibnamefont {Li}},\ }\bibfield  {title} {\bibinfo
  {title} {Extremely asymmetrical acoustic metasurface mirror at the
  exceptional point},\ }\href {https://doi.org/10.1103/PhysRevLett.123.214302}
  {\bibfield  {journal} {\bibinfo  {journal} {Phys. Rev. Lett.}\ }\textbf
  {\bibinfo {volume} {123}},\ \bibinfo {pages} {214302} (\bibinfo {year}
  {2019})}\BibitemShut {NoStop}%
\bibitem [{\citenamefont {Wiersig}(2014)}]{PhysRevLett.112.203901}%
  \BibitemOpen
  \bibfield  {author} {\bibinfo {author} {\bibfnamefont {J.}~\bibnamefont
  {Wiersig}},\ }\bibfield  {title} {\bibinfo {title} {Enhancing the sensitivity
  of frequency and energy splitting detection by using exceptional points:
  Application to microcavity sensors for single-particle detection},\ }\href
  {https://doi.org/10.1103/PhysRevLett.112.203901} {\bibfield  {journal}
  {\bibinfo  {journal} {Phys. Rev. Lett.}\ }\textbf {\bibinfo {volume} {112}},\
  \bibinfo {pages} {203901} (\bibinfo {year} {2014})}\BibitemShut {NoStop}%
\bibitem [{\citenamefont {Chen}\ \emph {et~al.}(2017)\citenamefont {Chen},
  \citenamefont {Kaya~\"{O}zdemir}, \citenamefont {Zhao}, \citenamefont
  {Wiersig},\ and\ \citenamefont {Yang}}]{Chen2017}%
  \BibitemOpen
  \bibfield  {author} {\bibinfo {author} {\bibfnamefont {W.}~\bibnamefont
  {Chen}}, \bibinfo {author} {\bibfnamefont {c.}~\bibnamefont
  {Kaya~\"{O}zdemir}}, \bibinfo {author} {\bibfnamefont {G.}~\bibnamefont
  {Zhao}}, \bibinfo {author} {\bibfnamefont {J.}~\bibnamefont {Wiersig}},\ and\
  \bibinfo {author} {\bibfnamefont {L.}~\bibnamefont {Yang}},\ }\bibfield
  {title} {\bibinfo {title} {Exceptional points enhance sensing in an optical
  microcavity},\ }\href {https://doi.org/10.1038/nature23281} {\bibfield
  {journal} {\bibinfo  {journal} {Nature (London)}\ }\textbf {\bibinfo {volume}
  {548}},\ \bibinfo {pages} {192} (\bibinfo {year} {2017})}\BibitemShut
  {NoStop}%
\bibitem [{\citenamefont {Hokmabadi}\ \emph {et~al.}(2019)\citenamefont
  {Hokmabadi}, \citenamefont {Schumer}, \citenamefont {Christodoulides},\ and\
  \citenamefont {Khajavikhan}}]{Hokmabadi2019}%
  \BibitemOpen
  \bibfield  {author} {\bibinfo {author} {\bibfnamefont {M.~P.}\ \bibnamefont
  {Hokmabadi}}, \bibinfo {author} {\bibfnamefont {A.}~\bibnamefont {Schumer}},
  \bibinfo {author} {\bibfnamefont {D.~N.}\ \bibnamefont {Christodoulides}},\
  and\ \bibinfo {author} {\bibfnamefont {M.}~\bibnamefont {Khajavikhan}},\
  }\bibfield  {title} {\bibinfo {title} {Non-hermitian ring laser gyroscopes
  with enhanced sagnac sensitivity},\ }\href
  {https://doi.org/10.1038/s41586-019-1780-4} {\bibfield  {journal} {\bibinfo
  {journal} {Nature (London)}\ }\textbf {\bibinfo {volume} {576}},\ \bibinfo
  {pages} {70} (\bibinfo {year} {2019})}\BibitemShut {NoStop}%
\bibitem [{\citenamefont {Lee}\ and\ \citenamefont
  {Chan}(2014)}]{PhysRevX.4.041001}%
  \BibitemOpen
  \bibfield  {author} {\bibinfo {author} {\bibfnamefont {T.~E.}\ \bibnamefont
  {Lee}}\ and\ \bibinfo {author} {\bibfnamefont {C.-K.}\ \bibnamefont {Chan}},\
  }\bibfield  {title} {\bibinfo {title} {Heralded magnetism in non-hermitian
  atomic systems},\ }\href {https://doi.org/10.1103/PhysRevX.4.041001}
  {\bibfield  {journal} {\bibinfo  {journal} {Phys. Rev. X}\ }\textbf {\bibinfo
  {volume} {4}},\ \bibinfo {pages} {041001} (\bibinfo {year}
  {2014})}\BibitemShut {NoStop}%
\bibitem [{\citenamefont {Kawabata}\ \emph {et~al.}(2017)\citenamefont
  {Kawabata}, \citenamefont {Ashida},\ and\ \citenamefont
  {Ueda}}]{PhysRevLett.119.190401}%
  \BibitemOpen
  \bibfield  {author} {\bibinfo {author} {\bibfnamefont {K.}~\bibnamefont
  {Kawabata}}, \bibinfo {author} {\bibfnamefont {Y.}~\bibnamefont {Ashida}},\
  and\ \bibinfo {author} {\bibfnamefont {M.}~\bibnamefont {Ueda}},\ }\bibfield
  {title} {\bibinfo {title} {Information retrieval and criticality in
  parity-time-symmetric systems},\ }\href
  {https://doi.org/10.1103/PhysRevLett.119.190401} {\bibfield  {journal}
  {\bibinfo  {journal} {Phys. Rev. Lett.}\ }\textbf {\bibinfo {volume} {119}},\
  \bibinfo {pages} {190401} (\bibinfo {year} {2017})}\BibitemShut {NoStop}%
\bibitem [{\citenamefont {Li}\ \emph {et~al.}(2019)\citenamefont {Li},
  \citenamefont {Harter}, \citenamefont {Liu}, \citenamefont {de~Melo},
  \citenamefont {Joglekar},\ and\ \citenamefont {Luo}}]{Li2019}%
  \BibitemOpen
  \bibfield  {author} {\bibinfo {author} {\bibfnamefont {J.}~\bibnamefont
  {Li}}, \bibinfo {author} {\bibfnamefont {A.~K.}\ \bibnamefont {Harter}},
  \bibinfo {author} {\bibfnamefont {J.}~\bibnamefont {Liu}}, \bibinfo {author}
  {\bibfnamefont {L.}~\bibnamefont {de~Melo}}, \bibinfo {author} {\bibfnamefont
  {Y.~N.}\ \bibnamefont {Joglekar}},\ and\ \bibinfo {author} {\bibfnamefont
  {L.}~\bibnamefont {Luo}},\ }\bibfield  {title} {\bibinfo {title} {Observation
  of parity-time symmetry breaking transitions in a dissipative floquet system
  of ultracold atoms},\ }\href {https://doi.org/10.1038/s41467-019-08596-1}
  {\bibfield  {journal} {\bibinfo  {journal} {Nat. Commun.}\ }\textbf {\bibinfo
  {volume} {10}},\ \bibinfo {pages} {855} (\bibinfo {year} {2019})}\BibitemShut
  {NoStop}%
\bibitem [{\citenamefont {D\'{o}ra}\ \emph {et~al.}(2019)\citenamefont
  {D\'{o}ra}, \citenamefont {Heyl},\ and\ \citenamefont {Moessner}}]{Dora2019}%
  \BibitemOpen
  \bibfield  {author} {\bibinfo {author} {\bibfnamefont {B.}~\bibnamefont
  {D\'{o}ra}}, \bibinfo {author} {\bibfnamefont {M.}~\bibnamefont {Heyl}},\
  and\ \bibinfo {author} {\bibfnamefont {R.}~\bibnamefont {Moessner}},\
  }\bibfield  {title} {\bibinfo {title} {The kibble-zurek mechanism at
  exceptional points},\ }\href {https://doi.org/10.1038/s41467-019-10048-9}
  {\bibfield  {journal} {\bibinfo  {journal} {Nat. Commun.}\ }\textbf {\bibinfo
  {volume} {10}},\ \bibinfo {pages} {2254} (\bibinfo {year}
  {2019})}\BibitemShut {NoStop}%
\bibitem [{\citenamefont {Xiao}\ \emph {et~al.}(2019)\citenamefont {Xiao},
  \citenamefont {Wang}, \citenamefont {Zhan}, \citenamefont {Bian},
  \citenamefont {Kawabata}, \citenamefont {Ueda}, \citenamefont {Yi},\ and\
  \citenamefont {Xue}}]{PhysRevLett.123.230401}%
  \BibitemOpen
  \bibfield  {author} {\bibinfo {author} {\bibfnamefont {L.}~\bibnamefont
  {Xiao}}, \bibinfo {author} {\bibfnamefont {K.}~\bibnamefont {Wang}}, \bibinfo
  {author} {\bibfnamefont {X.}~\bibnamefont {Zhan}}, \bibinfo {author}
  {\bibfnamefont {Z.}~\bibnamefont {Bian}}, \bibinfo {author} {\bibfnamefont
  {K.}~\bibnamefont {Kawabata}}, \bibinfo {author} {\bibfnamefont
  {M.}~\bibnamefont {Ueda}}, \bibinfo {author} {\bibfnamefont {W.}~\bibnamefont
  {Yi}},\ and\ \bibinfo {author} {\bibfnamefont {P.}~\bibnamefont {Xue}},\
  }\bibfield  {title} {\bibinfo {title} {Observation of critical phenomena in
  parity-time-symmetric quantum dynamics},\ }\href
  {https://doi.org/10.1103/PhysRevLett.123.230401} {\bibfield  {journal}
  {\bibinfo  {journal} {Phys. Rev. Lett.}\ }\textbf {\bibinfo {volume} {123}},\
  \bibinfo {pages} {230401} (\bibinfo {year} {2019})}\BibitemShut {NoStop}%
\bibitem [{\citenamefont {Wang}\ \emph {et~al.}(2021)\citenamefont {Wang},
  \citenamefont {Dutt}, \citenamefont {Wojcik},\ and\ \citenamefont
  {Fan}}]{Wang2021}%
  \BibitemOpen
  \bibfield  {author} {\bibinfo {author} {\bibfnamefont {K.}~\bibnamefont
  {Wang}}, \bibinfo {author} {\bibfnamefont {A.}~\bibnamefont {Dutt}}, \bibinfo
  {author} {\bibfnamefont {C.~C.}\ \bibnamefont {Wojcik}},\ and\ \bibinfo
  {author} {\bibfnamefont {S.}~\bibnamefont {Fan}},\ }\bibfield  {title}
  {\bibinfo {title} {Topological complex-energy braiding of non-hermitian
  bands},\ }\href {https://doi.org/10.1038/s41586-021-03848-x} {\bibfield
  {journal} {\bibinfo  {journal} {Nature (London)}\ }\textbf {\bibinfo {volume}
  {598}},\ \bibinfo {pages} {59} (\bibinfo {year} {2021})}\BibitemShut
  {NoStop}%
\bibitem [{\citenamefont {Hu}\ and\ \citenamefont
  {Zhao}(2021)}]{PhysRevLett.126.010401}%
  \BibitemOpen
  \bibfield  {author} {\bibinfo {author} {\bibfnamefont {H.}~\bibnamefont
  {Hu}}\ and\ \bibinfo {author} {\bibfnamefont {E.}~\bibnamefont {Zhao}},\
  }\bibfield  {title} {\bibinfo {title} {Knots and non-hermitian bloch bands},\
  }\href {https://doi.org/10.1103/PhysRevLett.126.010401} {\bibfield  {journal}
  {\bibinfo  {journal} {Phys. Rev. Lett.}\ }\textbf {\bibinfo {volume} {126}},\
  \bibinfo {pages} {010401} (\bibinfo {year} {2021})}\BibitemShut {NoStop}%
\bibitem [{\citenamefont {Zhang}\ \emph
  {et~al.}(2023{\natexlab{a}})\citenamefont {Zhang}, \citenamefont {Li},
  \citenamefont {Sun}, \citenamefont {Liu}, \citenamefont {Zhao}, \citenamefont
  {Feng}, \citenamefont {Fan},\ and\ \citenamefont
  {Qiu}}]{PhysRevLett.130.017201}%
  \BibitemOpen
  \bibfield  {author} {\bibinfo {author} {\bibfnamefont {Q.}~\bibnamefont
  {Zhang}}, \bibinfo {author} {\bibfnamefont {Y.}~\bibnamefont {Li}}, \bibinfo
  {author} {\bibfnamefont {H.}~\bibnamefont {Sun}}, \bibinfo {author}
  {\bibfnamefont {X.}~\bibnamefont {Liu}}, \bibinfo {author} {\bibfnamefont
  {L.}~\bibnamefont {Zhao}}, \bibinfo {author} {\bibfnamefont {X.}~\bibnamefont
  {Feng}}, \bibinfo {author} {\bibfnamefont {X.}~\bibnamefont {Fan}},\ and\
  \bibinfo {author} {\bibfnamefont {C.}~\bibnamefont {Qiu}},\ }\bibfield
  {title} {\bibinfo {title} {Observation of acoustic non-hermitian bloch braids
  and associated topological phase transitions},\ }\href
  {https://doi.org/10.1103/PhysRevLett.130.017201} {\bibfield  {journal}
  {\bibinfo  {journal} {Phys. Rev. Lett.}\ }\textbf {\bibinfo {volume} {130}},\
  \bibinfo {pages} {017201} (\bibinfo {year} {2023}{\natexlab{a}})}\BibitemShut
  {NoStop}%
\bibitem [{\citenamefont {Cao}\ \emph {et~al.}(2023)\citenamefont {Cao},
  \citenamefont {Li}, \citenamefont {Zhao}, \citenamefont {Guo}, \citenamefont
  {Qi}, \citenamefont {Chang}, \citenamefont {Zhou}, \citenamefont {Xu},\ and\
  \citenamefont {Duan}}]{PhysRevLett.130.163001}%
  \BibitemOpen
  \bibfield  {author} {\bibinfo {author} {\bibfnamefont {M.-M.}\ \bibnamefont
  {Cao}}, \bibinfo {author} {\bibfnamefont {K.}~\bibnamefont {Li}}, \bibinfo
  {author} {\bibfnamefont {W.-D.}\ \bibnamefont {Zhao}}, \bibinfo {author}
  {\bibfnamefont {W.-X.}\ \bibnamefont {Guo}}, \bibinfo {author} {\bibfnamefont
  {B.-X.}\ \bibnamefont {Qi}}, \bibinfo {author} {\bibfnamefont {X.-Y.}\
  \bibnamefont {Chang}}, \bibinfo {author} {\bibfnamefont {Z.-C.}\ \bibnamefont
  {Zhou}}, \bibinfo {author} {\bibfnamefont {Y.}~\bibnamefont {Xu}},\ and\
  \bibinfo {author} {\bibfnamefont {L.-M.}\ \bibnamefont {Duan}},\ }\bibfield
  {title} {\bibinfo {title} {Probing complex-energy topology via non-hermitian
  absorption spectroscopy in a trapped ion simulator},\ }\href
  {https://doi.org/10.1103/PhysRevLett.130.163001} {\bibfield  {journal}
  {\bibinfo  {journal} {Phys. Rev. Lett.}\ }\textbf {\bibinfo {volume} {130}},\
  \bibinfo {pages} {163001} (\bibinfo {year} {2023})}\BibitemShut {NoStop}%
\bibitem [{\citenamefont {Hu}\ \emph {et~al.}(2022)\citenamefont {Hu},
  \citenamefont {Sun},\ and\ \citenamefont {Chen}}]{PhysRevResearch.4.L022064}%
  \BibitemOpen
  \bibfield  {author} {\bibinfo {author} {\bibfnamefont {H.}~\bibnamefont
  {Hu}}, \bibinfo {author} {\bibfnamefont {S.}~\bibnamefont {Sun}},\ and\
  \bibinfo {author} {\bibfnamefont {S.}~\bibnamefont {Chen}},\ }\bibfield
  {title} {\bibinfo {title} {Knot topology of exceptional point and
  non-hermitian no-go theorem},\ }\href
  {https://doi.org/10.1103/PhysRevResearch.4.L022064} {\bibfield  {journal}
  {\bibinfo  {journal} {Phys. Rev. Res.}\ }\textbf {\bibinfo {volume} {4}},\
  \bibinfo {pages} {L022064} (\bibinfo {year} {2022})}\BibitemShut {NoStop}%
\bibitem [{\citenamefont {Qi}\ \emph {et~al.}(2024)\citenamefont {Qi},
  \citenamefont {Li}, \citenamefont {Wang}, \citenamefont {Li}, \citenamefont
  {Li}, \citenamefont {Wang}, \citenamefont {Hu},\ and\ \citenamefont
  {Gong}}]{PhysRevLett.132.243802}%
  \BibitemOpen
  \bibfield  {author} {\bibinfo {author} {\bibfnamefont {H.}~\bibnamefont
  {Qi}}, \bibinfo {author} {\bibfnamefont {Y.}~\bibnamefont {Li}}, \bibinfo
  {author} {\bibfnamefont {X.}~\bibnamefont {Wang}}, \bibinfo {author}
  {\bibfnamefont {Y.}~\bibnamefont {Li}}, \bibinfo {author} {\bibfnamefont
  {X.}~\bibnamefont {Li}}, \bibinfo {author} {\bibfnamefont {X.}~\bibnamefont
  {Wang}}, \bibinfo {author} {\bibfnamefont {X.}~\bibnamefont {Hu}},\ and\
  \bibinfo {author} {\bibfnamefont {Q.}~\bibnamefont {Gong}},\ }\bibfield
  {title} {\bibinfo {title} {Dynamically encircling exceptional points in
  different riemann sheets for orbital angular momentum topological charge
  conversion},\ }\href {https://doi.org/10.1103/PhysRevLett.132.243802}
  {\bibfield  {journal} {\bibinfo  {journal} {Phys. Rev. Lett.}\ }\textbf
  {\bibinfo {volume} {132}},\ \bibinfo {pages} {243802} (\bibinfo {year}
  {2024})}\BibitemShut {NoStop}%
\bibitem [{\citenamefont {Doppler}\ \emph {et~al.}(2016)\citenamefont
  {Doppler}, \citenamefont {Mailybaev}, \citenamefont {B\''{o}hm},
  \citenamefont {Kuhl}, \citenamefont {Girschik}, \citenamefont {Libisch},
  \citenamefont {Milburn}, \citenamefont {Rabl}, \citenamefont {Moiseyev},\
  and\ \citenamefont {Rotter}}]{Doppler2016}%
  \BibitemOpen
  \bibfield  {author} {\bibinfo {author} {\bibfnamefont {J.}~\bibnamefont
  {Doppler}}, \bibinfo {author} {\bibfnamefont {A.~A.}\ \bibnamefont
  {Mailybaev}}, \bibinfo {author} {\bibfnamefont {J.}~\bibnamefont
  {B\''{o}hm}}, \bibinfo {author} {\bibfnamefont {U.}~\bibnamefont {Kuhl}},
  \bibinfo {author} {\bibfnamefont {A.}~\bibnamefont {Girschik}}, \bibinfo
  {author} {\bibfnamefont {F.}~\bibnamefont {Libisch}}, \bibinfo {author}
  {\bibfnamefont {T.~J.}\ \bibnamefont {Milburn}}, \bibinfo {author}
  {\bibfnamefont {P.}~\bibnamefont {Rabl}}, \bibinfo {author} {\bibfnamefont
  {N.}~\bibnamefont {Moiseyev}},\ and\ \bibinfo {author} {\bibfnamefont
  {S.}~\bibnamefont {Rotter}},\ }\bibfield  {title} {\bibinfo {title}
  {Dynamically encircling an exceptional point for asymmetric mode switching},\
  }\href {https://doi.org/10.1038/nature18605} {\bibfield  {journal} {\bibinfo
  {journal} {Nature (London)}\ }\textbf {\bibinfo {volume} {537}},\ \bibinfo
  {pages} {76} (\bibinfo {year} {2016})}\BibitemShut {NoStop}%
\bibitem [{\citenamefont {Xu}\ \emph {et~al.}(2016)\citenamefont {Xu},
  \citenamefont {Mason}, \citenamefont {Jiang},\ and\ \citenamefont
  {Harris}}]{Xu2016}%
  \BibitemOpen
  \bibfield  {author} {\bibinfo {author} {\bibfnamefont {H.}~\bibnamefont
  {Xu}}, \bibinfo {author} {\bibfnamefont {D.}~\bibnamefont {Mason}}, \bibinfo
  {author} {\bibfnamefont {L.}~\bibnamefont {Jiang}},\ and\ \bibinfo {author}
  {\bibfnamefont {J.~G.~E.}\ \bibnamefont {Harris}},\ }\bibfield  {title}
  {\bibinfo {title} {Topological energy transfer in an optomechanical system
  with exceptional points},\ }\href {https://doi.org/10.1038/nature18604}
  {\bibfield  {journal} {\bibinfo  {journal} {Nature (London)}\ }\textbf
  {\bibinfo {volume} {537}},\ \bibinfo {pages} {80} (\bibinfo {year}
  {2016})}\BibitemShut {NoStop}%
\bibitem [{\citenamefont {Yoon}\ \emph {et~al.}(2018)\citenamefont {Yoon},
  \citenamefont {Choi}, \citenamefont {Hahn}, \citenamefont {Kim},
  \citenamefont {Song}, \citenamefont {Yang}, \citenamefont {Lee},
  \citenamefont {Kim}, \citenamefont {Lee}, \citenamefont {Shin}, \citenamefont
  {Lee},\ and\ \citenamefont {Berini}}]{Yoon2018}%
  \BibitemOpen
  \bibfield  {author} {\bibinfo {author} {\bibfnamefont {J.~W.}\ \bibnamefont
  {Yoon}}, \bibinfo {author} {\bibfnamefont {Y.}~\bibnamefont {Choi}}, \bibinfo
  {author} {\bibfnamefont {C.}~\bibnamefont {Hahn}}, \bibinfo {author}
  {\bibfnamefont {G.}~\bibnamefont {Kim}}, \bibinfo {author} {\bibfnamefont
  {S.~H.}\ \bibnamefont {Song}}, \bibinfo {author} {\bibfnamefont {K.-Y.}\
  \bibnamefont {Yang}}, \bibinfo {author} {\bibfnamefont {J.~Y.}\ \bibnamefont
  {Lee}}, \bibinfo {author} {\bibfnamefont {Y.}~\bibnamefont {Kim}}, \bibinfo
  {author} {\bibfnamefont {C.~S.}\ \bibnamefont {Lee}}, \bibinfo {author}
  {\bibfnamefont {J.~K.}\ \bibnamefont {Shin}}, \bibinfo {author}
  {\bibfnamefont {H.-S.}\ \bibnamefont {Lee}},\ and\ \bibinfo {author}
  {\bibfnamefont {P.}~\bibnamefont {Berini}},\ }\bibfield  {title} {\bibinfo
  {title} {Time-asymmetric loop around an exceptional point over the full
  optical communications band},\ }\href
  {https://doi.org/10.1038/s41586-018-0523-2} {\bibfield  {journal} {\bibinfo
  {journal} {Nature (London)}\ }\textbf {\bibinfo {volume} {7725}},\ \bibinfo
  {pages} {86} (\bibinfo {year} {2018})}\BibitemShut {NoStop}%
\bibitem [{\citenamefont {Schumer}\ \emph {et~al.}(2022)\citenamefont
  {Schumer}, \citenamefont {Liu}, \citenamefont {Leshin}, \citenamefont {Ding},
  \citenamefont {Alahmadi}, \citenamefont {Hassan}, \citenamefont {Nasari},
  \citenamefont {Rotter}, \citenamefont {Christodoulides}, \citenamefont
  {LiKamWa},\ and\ \citenamefont {Khajavikhan}}]{doi:10.1126/science.abl6571}%
  \BibitemOpen
  \bibfield  {author} {\bibinfo {author} {\bibfnamefont {A.}~\bibnamefont
  {Schumer}}, \bibinfo {author} {\bibfnamefont {Y.~G.~N.}\ \bibnamefont {Liu}},
  \bibinfo {author} {\bibfnamefont {J.}~\bibnamefont {Leshin}}, \bibinfo
  {author} {\bibfnamefont {L.}~\bibnamefont {Ding}}, \bibinfo {author}
  {\bibfnamefont {Y.}~\bibnamefont {Alahmadi}}, \bibinfo {author}
  {\bibfnamefont {A.~U.}\ \bibnamefont {Hassan}}, \bibinfo {author}
  {\bibfnamefont {H.}~\bibnamefont {Nasari}}, \bibinfo {author} {\bibfnamefont
  {S.}~\bibnamefont {Rotter}}, \bibinfo {author} {\bibfnamefont {D.~N.}\
  \bibnamefont {Christodoulides}}, \bibinfo {author} {\bibfnamefont
  {P.}~\bibnamefont {LiKamWa}},\ and\ \bibinfo {author} {\bibfnamefont
  {M.}~\bibnamefont {Khajavikhan}},\ }\bibfield  {title} {\bibinfo {title}
  {Topological modes in a laser cavity through exceptional state transfer},\
  }\href {https://doi.org/10.1126/science.abl6571} {\bibfield  {journal}
  {\bibinfo  {journal} {Science}\ }\textbf {\bibinfo {volume} {375}},\ \bibinfo
  {pages} {884} (\bibinfo {year} {2022})}\BibitemShut {NoStop}%
\bibitem [{\citenamefont {Liu}\ \emph {et~al.}(2021)\citenamefont {Liu},
  \citenamefont {Wu}, \citenamefont {Duan}, \citenamefont {Rong},\ and\
  \citenamefont {Du}}]{PhysRevLett.126.170506}%
  \BibitemOpen
  \bibfield  {author} {\bibinfo {author} {\bibfnamefont {W.}~\bibnamefont
  {Liu}}, \bibinfo {author} {\bibfnamefont {Y.}~\bibnamefont {Wu}}, \bibinfo
  {author} {\bibfnamefont {C.-K.}\ \bibnamefont {Duan}}, \bibinfo {author}
  {\bibfnamefont {X.}~\bibnamefont {Rong}},\ and\ \bibinfo {author}
  {\bibfnamefont {J.}~\bibnamefont {Du}},\ }\bibfield  {title} {\bibinfo
  {title} {Dynamically encircling an exceptional point in a real quantum
  system},\ }\href {https://doi.org/10.1103/PhysRevLett.126.170506} {\bibfield
  {journal} {\bibinfo  {journal} {Phys. Rev. Lett.}\ }\textbf {\bibinfo
  {volume} {126}},\ \bibinfo {pages} {170506} (\bibinfo {year}
  {2021})}\BibitemShut {NoStop}%
\bibitem [{\citenamefont {Guo}\ \emph {et~al.}(2023)\citenamefont {Guo},
  \citenamefont {Chen}, \citenamefont {Ding},\ and\ \citenamefont
  {Hu}}]{PhysRevLett.130.157201}%
  \BibitemOpen
  \bibfield  {author} {\bibinfo {author} {\bibfnamefont {C.-X.}\ \bibnamefont
  {Guo}}, \bibinfo {author} {\bibfnamefont {S.}~\bibnamefont {Chen}}, \bibinfo
  {author} {\bibfnamefont {K.}~\bibnamefont {Ding}},\ and\ \bibinfo {author}
  {\bibfnamefont {H.}~\bibnamefont {Hu}},\ }\bibfield  {title} {\bibinfo
  {title} {Exceptional non-abelian topology in multiband non-hermitian
  systems},\ }\href {https://doi.org/10.1103/PhysRevLett.130.157201} {\bibfield
   {journal} {\bibinfo  {journal} {Phys. Rev. Lett.}\ }\textbf {\bibinfo
  {volume} {130}},\ \bibinfo {pages} {157201} (\bibinfo {year}
  {2023})}\BibitemShut {NoStop}%
\bibitem [{\citenamefont {Demange}\ and\ \citenamefont
  {Graefe}(2011)}]{Demange_2012}%
  \BibitemOpen
  \bibfield  {author} {\bibinfo {author} {\bibfnamefont {G.}~\bibnamefont
  {Demange}}\ and\ \bibinfo {author} {\bibfnamefont {E.-M.}\ \bibnamefont
  {Graefe}},\ }\bibfield  {title} {\bibinfo {title} {Signatures of three
  coalescing eigenfunctions},\ }\href
  {https://doi.org/10.1088/1751-8113/45/2/025303} {\bibfield  {journal}
  {\bibinfo  {journal} {J. Phys. A}\ }\textbf {\bibinfo {volume} {45}},\
  \bibinfo {pages} {025303} (\bibinfo {year} {2011})}\BibitemShut {NoStop}%
\bibitem [{\citenamefont {Mandal}\ and\ \citenamefont
  {Bergholtz}(2021)}]{PhysRevLett.127.186601}%
  \BibitemOpen
  \bibfield  {author} {\bibinfo {author} {\bibfnamefont {I.}~\bibnamefont
  {Mandal}}\ and\ \bibinfo {author} {\bibfnamefont {E.~J.}\ \bibnamefont
  {Bergholtz}},\ }\bibfield  {title} {\bibinfo {title} {Symmetry and
  higher-order exceptional points},\ }\href
  {https://doi.org/10.1103/PhysRevLett.127.186601} {\bibfield  {journal}
  {\bibinfo  {journal} {Phys. Rev. Lett.}\ }\textbf {\bibinfo {volume} {127}},\
  \bibinfo {pages} {186601} (\bibinfo {year} {2021})}\BibitemShut {NoStop}%
\bibitem [{\citenamefont {Delplace}\ \emph {et~al.}(2021)\citenamefont
  {Delplace}, \citenamefont {Yoshida},\ and\ \citenamefont
  {Hatsugai}}]{PhysRevLett.127.186602}%
  \BibitemOpen
  \bibfield  {author} {\bibinfo {author} {\bibfnamefont {P.}~\bibnamefont
  {Delplace}}, \bibinfo {author} {\bibfnamefont {T.}~\bibnamefont {Yoshida}},\
  and\ \bibinfo {author} {\bibfnamefont {Y.}~\bibnamefont {Hatsugai}},\
  }\bibfield  {title} {\bibinfo {title} {Symmetry-protected multifold
  exceptional points and their topological characterization},\ }\href
  {https://doi.org/10.1103/PhysRevLett.127.186602} {\bibfield  {journal}
  {\bibinfo  {journal} {Phys. Rev. Lett.}\ }\textbf {\bibinfo {volume} {127}},\
  \bibinfo {pages} {186602} (\bibinfo {year} {2021})}\BibitemShut {NoStop}%
\bibitem [{\citenamefont {Bai}\ \emph {et~al.}(2023)\citenamefont {Bai},
  \citenamefont {Li}, \citenamefont {Liu}, \citenamefont {Fang}, \citenamefont
  {Wan},\ and\ \citenamefont {Xiao}}]{PhysRevLett.130.266901}%
  \BibitemOpen
  \bibfield  {author} {\bibinfo {author} {\bibfnamefont {K.}~\bibnamefont
  {Bai}}, \bibinfo {author} {\bibfnamefont {J.-Z.}\ \bibnamefont {Li}},
  \bibinfo {author} {\bibfnamefont {T.-R.}\ \bibnamefont {Liu}}, \bibinfo
  {author} {\bibfnamefont {L.}~\bibnamefont {Fang}}, \bibinfo {author}
  {\bibfnamefont {D.}~\bibnamefont {Wan}},\ and\ \bibinfo {author}
  {\bibfnamefont {M.}~\bibnamefont {Xiao}},\ }\bibfield  {title} {\bibinfo
  {title} {Nonlinear exceptional points with a complete basis in dynamics},\
  }\href {https://doi.org/10.1103/PhysRevLett.130.266901} {\bibfield  {journal}
  {\bibinfo  {journal} {Phys. Rev. Lett.}\ }\textbf {\bibinfo {volume} {130}},\
  \bibinfo {pages} {266901} (\bibinfo {year} {2023})}\BibitemShut {NoStop}%
\bibitem [{\citenamefont {Wang}\ \emph {et~al.}(2023)\citenamefont {Wang},
  \citenamefont {Xiao}, \citenamefont {Lin}, \citenamefont {Yi}, \citenamefont
  {Bergholtz},\ and\ \citenamefont {Xue}}]{doi:10.1126/sciadv.adi0732}%
  \BibitemOpen
  \bibfield  {author} {\bibinfo {author} {\bibfnamefont {K.}~\bibnamefont
  {Wang}}, \bibinfo {author} {\bibfnamefont {L.}~\bibnamefont {Xiao}}, \bibinfo
  {author} {\bibfnamefont {H.}~\bibnamefont {Lin}}, \bibinfo {author}
  {\bibfnamefont {W.}~\bibnamefont {Yi}}, \bibinfo {author} {\bibfnamefont
  {E.~J.}\ \bibnamefont {Bergholtz}},\ and\ \bibinfo {author} {\bibfnamefont
  {P.}~\bibnamefont {Xue}},\ }\bibfield  {title} {\bibinfo {title}
  {Experimental simulation of symmetry-protected higher-order exceptional
  points with single photons},\ }\href {https://doi.org/10.1126/sciadv.adi0732}
  {\bibfield  {journal} {\bibinfo  {journal} {Sci. Adv.}\ }\textbf {\bibinfo
  {volume} {9}},\ \bibinfo {pages} {eadi0732} (\bibinfo {year}
  {2023})}\BibitemShut {NoStop}%
\bibitem [{\citenamefont {Li}\ \emph {et~al.}(2023)\citenamefont {Li},
  \citenamefont {Chen}, \citenamefont {Abbasi}, \citenamefont {Murch},\ and\
  \citenamefont {Whaley}}]{PhysRevLett.131.100202}%
  \BibitemOpen
  \bibfield  {author} {\bibinfo {author} {\bibfnamefont {Z.-Z.}\ \bibnamefont
  {Li}}, \bibinfo {author} {\bibfnamefont {W.}~\bibnamefont {Chen}}, \bibinfo
  {author} {\bibfnamefont {M.}~\bibnamefont {Abbasi}}, \bibinfo {author}
  {\bibfnamefont {K.~W.}\ \bibnamefont {Murch}},\ and\ \bibinfo {author}
  {\bibfnamefont {K.~B.}\ \bibnamefont {Whaley}},\ }\bibfield  {title}
  {\bibinfo {title} {Speeding up entanglement generation by proximity to
  higher-order exceptional points},\ }\href
  {https://doi.org/10.1103/PhysRevLett.131.100202} {\bibfield  {journal}
  {\bibinfo  {journal} {Phys. Rev. Lett.}\ }\textbf {\bibinfo {volume} {131}},\
  \bibinfo {pages} {100202} (\bibinfo {year} {2023})}\BibitemShut {NoStop}%
\bibitem [{\citenamefont {Xu}\ \emph {et~al.}(2017)\citenamefont {Xu},
  \citenamefont {Wang},\ and\ \citenamefont {Duan}}]{PhysRevLett.118.045701}%
  \BibitemOpen
  \bibfield  {author} {\bibinfo {author} {\bibfnamefont {Y.}~\bibnamefont
  {Xu}}, \bibinfo {author} {\bibfnamefont {S.-T.}\ \bibnamefont {Wang}},\ and\
  \bibinfo {author} {\bibfnamefont {L.-M.}\ \bibnamefont {Duan}},\ }\bibfield
  {title} {\bibinfo {title} {Weyl exceptional rings in a three-dimensional
  dissipative cold atomic gas},\ }\href
  {https://doi.org/10.1103/PhysRevLett.118.045701} {\bibfield  {journal}
  {\bibinfo  {journal} {Phys. Rev. Lett.}\ }\textbf {\bibinfo {volume} {118}},\
  \bibinfo {pages} {045701} (\bibinfo {year} {2017})}\BibitemShut {NoStop}%
\bibitem [{\citenamefont {Cerjan}\ \emph {et~al.}(2019)\citenamefont {Cerjan},
  \citenamefont {Huang}, \citenamefont {Wang}, \citenamefont {Chen},
  \citenamefont {Chong},\ and\ \citenamefont {Rechtsman}}]{Cerjan2019}%
  \BibitemOpen
  \bibfield  {author} {\bibinfo {author} {\bibfnamefont {A.}~\bibnamefont
  {Cerjan}}, \bibinfo {author} {\bibfnamefont {S.}~\bibnamefont {Huang}},
  \bibinfo {author} {\bibfnamefont {M.}~\bibnamefont {Wang}}, \bibinfo {author}
  {\bibfnamefont {K.~P.}\ \bibnamefont {Chen}}, \bibinfo {author}
  {\bibfnamefont {Y.}~\bibnamefont {Chong}},\ and\ \bibinfo {author}
  {\bibfnamefont {M.~C.}\ \bibnamefont {Rechtsman}},\ }\bibfield  {title}
  {\bibinfo {title} {Experimental realization of a weyl exceptional ring},\
  }\href {https://doi.org/10.1038/s41566-019-0453-z} {\bibfield  {journal}
  {\bibinfo  {journal} {Nat. Photonics}\ }\textbf {\bibinfo {volume} {13}},\
  \bibinfo {pages} {623} (\bibinfo {year} {2019})}\BibitemShut {NoStop}%
\bibitem [{\citenamefont {Liu}\ \emph {et~al.}(2022)\citenamefont {Liu},
  \citenamefont {Li}, \citenamefont {Chen}, \citenamefont {Tang}, \citenamefont
  {Chen}, \citenamefont {Liang}, \citenamefont {Ma},\ and\ \citenamefont
  {Cheng}}]{PhysRevLett.129.084301}%
  \BibitemOpen
  \bibfield  {author} {\bibinfo {author} {\bibfnamefont {J.-j.}\ \bibnamefont
  {Liu}}, \bibinfo {author} {\bibfnamefont {Z.-w.}\ \bibnamefont {Li}},
  \bibinfo {author} {\bibfnamefont {Z.-G.}\ \bibnamefont {Chen}}, \bibinfo
  {author} {\bibfnamefont {W.}~\bibnamefont {Tang}}, \bibinfo {author}
  {\bibfnamefont {A.}~\bibnamefont {Chen}}, \bibinfo {author} {\bibfnamefont
  {B.}~\bibnamefont {Liang}}, \bibinfo {author} {\bibfnamefont
  {G.}~\bibnamefont {Ma}},\ and\ \bibinfo {author} {\bibfnamefont {J.-C.}\
  \bibnamefont {Cheng}},\ }\bibfield  {title} {\bibinfo {title} {Experimental
  realization of weyl exceptional rings in a synthetic three-dimensional
  non-hermitian phononic crystal},\ }\href
  {https://doi.org/10.1103/PhysRevLett.129.084301} {\bibfield  {journal}
  {\bibinfo  {journal} {Phys. Rev. Lett.}\ }\textbf {\bibinfo {volume} {129}},\
  \bibinfo {pages} {084301} (\bibinfo {year} {2022})}\BibitemShut {NoStop}%
\bibitem [{\citenamefont {Xu}\ \emph {et~al.}(2022)\citenamefont {Xu},
  \citenamefont {Li}, \citenamefont {Zhou}, \citenamefont {Li}, \citenamefont
  {Li}, \citenamefont {Fan}, \citenamefont {Zhang}, \citenamefont
  {Christodoulides},\ and\ \citenamefont {Qiu}}]{doi:10.1073/pnas.2110018119}%
  \BibitemOpen
  \bibfield  {author} {\bibinfo {author} {\bibfnamefont {G.}~\bibnamefont
  {Xu}}, \bibinfo {author} {\bibfnamefont {W.}~\bibnamefont {Li}}, \bibinfo
  {author} {\bibfnamefont {X.}~\bibnamefont {Zhou}}, \bibinfo {author}
  {\bibfnamefont {H.}~\bibnamefont {Li}}, \bibinfo {author} {\bibfnamefont
  {Y.}~\bibnamefont {Li}}, \bibinfo {author} {\bibfnamefont {S.}~\bibnamefont
  {Fan}}, \bibinfo {author} {\bibfnamefont {S.}~\bibnamefont {Zhang}}, \bibinfo
  {author} {\bibfnamefont {D.~N.}\ \bibnamefont {Christodoulides}},\ and\
  \bibinfo {author} {\bibfnamefont {C.-W.}\ \bibnamefont {Qiu}},\ }\bibfield
  {title} {\bibinfo {title} {Observation of weyl exceptional rings in thermal
  diffusion},\ }\href {https://doi.org/10.1073/pnas.2110018119} {\bibfield
  {journal} {\bibinfo  {journal} {Proc. Natl. Acad. Sci. U.S.A}\ }\textbf
  {\bibinfo {volume} {119}},\ \bibinfo {pages} {e2110018119} (\bibinfo {year}
  {2022})}\BibitemShut {NoStop}%
\bibitem [{\citenamefont {Wu}\ \emph {et~al.}(2024)\citenamefont {Wu},
  \citenamefont {Wang}, \citenamefont {Ye}, \citenamefont {Liu}, \citenamefont
  {Niu}, \citenamefont {Duan}, \citenamefont {Wang}, \citenamefont {Rong},\
  and\ \citenamefont {Du}}]{Wu2024}%
  \BibitemOpen
  \bibfield  {author} {\bibinfo {author} {\bibfnamefont {Y.}~\bibnamefont
  {Wu}}, \bibinfo {author} {\bibfnamefont {Y.}~\bibnamefont {Wang}}, \bibinfo
  {author} {\bibfnamefont {X.}~\bibnamefont {Ye}}, \bibinfo {author}
  {\bibfnamefont {W.}~\bibnamefont {Liu}}, \bibinfo {author} {\bibfnamefont
  {Z.}~\bibnamefont {Niu}}, \bibinfo {author} {\bibfnamefont {C.-K.}\
  \bibnamefont {Duan}}, \bibinfo {author} {\bibfnamefont {Y.}~\bibnamefont
  {Wang}}, \bibinfo {author} {\bibfnamefont {X.}~\bibnamefont {Rong}},\ and\
  \bibinfo {author} {\bibfnamefont {J.}~\bibnamefont {Du}},\ }\bibfield
  {title} {\bibinfo {title} {Third-order exceptional line in a nitrogen-vacancy
  spin system},\ }\href {https://doi.org/10.1038/s41565-023-01583-0} {\bibfield
   {journal} {\bibinfo  {journal} {Nat. Nanotechnol.}\ } (\bibinfo {year}
  {2024})}\BibitemShut {NoStop}%
\bibitem [{\citenamefont {Tang}\ \emph {et~al.}(2020)\citenamefont {Tang},
  \citenamefont {Jiang}, \citenamefont {Ding}, \citenamefont {Xiao},
  \citenamefont {Zhang}, \citenamefont {Chan},\ and\ \citenamefont
  {Ma}}]{doi:10.1126/science.abd8872}%
  \BibitemOpen
  \bibfield  {author} {\bibinfo {author} {\bibfnamefont {W.}~\bibnamefont
  {Tang}}, \bibinfo {author} {\bibfnamefont {X.}~\bibnamefont {Jiang}},
  \bibinfo {author} {\bibfnamefont {K.}~\bibnamefont {Ding}}, \bibinfo {author}
  {\bibfnamefont {Y.-X.}\ \bibnamefont {Xiao}}, \bibinfo {author}
  {\bibfnamefont {Z.-Q.}\ \bibnamefont {Zhang}}, \bibinfo {author}
  {\bibfnamefont {C.~T.}\ \bibnamefont {Chan}},\ and\ \bibinfo {author}
  {\bibfnamefont {G.}~\bibnamefont {Ma}},\ }\bibfield  {title} {\bibinfo
  {title} {Exceptional nexus with a hybrid topological invariant},\ }\href
  {https://doi.org/10.1126/science.abd8872} {\bibfield  {journal} {\bibinfo
  {journal} {Science}\ }\textbf {\bibinfo {volume} {370}},\ \bibinfo {pages}
  {1077} (\bibinfo {year} {2020})}\BibitemShut {NoStop}%
\bibitem [{\citenamefont {Wang}\ \emph {et~al.}(2024)\citenamefont {Wang},
  \citenamefont {Li}, \citenamefont {Xie}, \citenamefont {Ding}, \citenamefont
  {Ji}, \citenamefont {Xiao}, \citenamefont {Jia}, \citenamefont {Yan},
  \citenamefont {Hu},\ and\ \citenamefont {Zhao}}]{PhysRevLett.132.253401}%
  \BibitemOpen
  \bibfield  {author} {\bibinfo {author} {\bibfnamefont {C.}~\bibnamefont
  {Wang}}, \bibinfo {author} {\bibfnamefont {N.}~\bibnamefont {Li}}, \bibinfo
  {author} {\bibfnamefont {J.}~\bibnamefont {Xie}}, \bibinfo {author}
  {\bibfnamefont {C.}~\bibnamefont {Ding}}, \bibinfo {author} {\bibfnamefont
  {Z.}~\bibnamefont {Ji}}, \bibinfo {author} {\bibfnamefont {L.}~\bibnamefont
  {Xiao}}, \bibinfo {author} {\bibfnamefont {S.}~\bibnamefont {Jia}}, \bibinfo
  {author} {\bibfnamefont {B.}~\bibnamefont {Yan}}, \bibinfo {author}
  {\bibfnamefont {Y.}~\bibnamefont {Hu}},\ and\ \bibinfo {author}
  {\bibfnamefont {Y.}~\bibnamefont {Zhao}},\ }\bibfield  {title} {\bibinfo
  {title} {Exceptional nexus in bose-einstein condensates with collective
  dissipation},\ }\href {https://doi.org/10.1103/PhysRevLett.132.253401}
  {\bibfield  {journal} {\bibinfo  {journal} {Phys. Rev. Lett.}\ }\textbf
  {\bibinfo {volume} {132}},\ \bibinfo {pages} {253401} (\bibinfo {year}
  {2024})}\BibitemShut {NoStop}%
\bibitem [{\citenamefont {Hu}\ \emph {et~al.}(2023)\citenamefont {Hu},
  \citenamefont {Zhang}, \citenamefont {Wang}, \citenamefont {Ouyang},
  \citenamefont {Zhu}, \citenamefont {Jia},\ and\ \citenamefont
  {Chan}}]{Hu2023}%
  \BibitemOpen
  \bibfield  {author} {\bibinfo {author} {\bibfnamefont {J.}~\bibnamefont
  {Hu}}, \bibinfo {author} {\bibfnamefont {R.-Y.}\ \bibnamefont {Zhang}},
  \bibinfo {author} {\bibfnamefont {Y.}~\bibnamefont {Wang}}, \bibinfo {author}
  {\bibfnamefont {X.}~\bibnamefont {Ouyang}}, \bibinfo {author} {\bibfnamefont
  {Y.}~\bibnamefont {Zhu}}, \bibinfo {author} {\bibfnamefont {H.}~\bibnamefont
  {Jia}},\ and\ \bibinfo {author} {\bibfnamefont {C.~T.}\ \bibnamefont
  {Chan}},\ }\bibfield  {title} {\bibinfo {title} {Non-hermitian swallowtail
  catastrophe revealing transitions among diverse topological singularities},\
  }\href {https://doi.org/10.1038/s41567-023-02048-w} {\bibfield  {journal}
  {\bibinfo  {journal} {Nat. Phys.}\ }\textbf {\bibinfo {volume} {19}},\
  \bibinfo {pages} {1098} (\bibinfo {year} {2023})}\BibitemShut {NoStop}%
\bibitem [{\citenamefont {Ding}\ \emph {et~al.}(2022)\citenamefont {Ding},
  \citenamefont {Fang},\ and\ \citenamefont {Ma}}]{Ding2022}%
  \BibitemOpen
  \bibfield  {author} {\bibinfo {author} {\bibfnamefont {K.}~\bibnamefont
  {Ding}}, \bibinfo {author} {\bibfnamefont {C.}~\bibnamefont {Fang}},\ and\
  \bibinfo {author} {\bibfnamefont {G.}~\bibnamefont {Ma}},\ }\bibfield
  {title} {\bibinfo {title} {Non-hermitian topology and exceptional-point
  geometries},\ }\href {https://doi.org/10.1038/s42254-022-00516-5} {\bibfield
  {journal} {\bibinfo  {journal} {Nat. Rev. Phy.}\ }\textbf {\bibinfo {volume}
  {4}},\ \bibinfo {pages} {745} (\bibinfo {year} {2022})}\BibitemShut {NoStop}%
\bibitem [{\citenamefont {Jangjan}\ \emph {et~al.}(2024)\citenamefont
  {Jangjan}, \citenamefont {Li}, \citenamefont {Foa~Torres},\ and\
  \citenamefont {Hosseini}}]{PhysRevB.109.205142}%
  \BibitemOpen
  \bibfield  {author} {\bibinfo {author} {\bibfnamefont {M.}~\bibnamefont
  {Jangjan}}, \bibinfo {author} {\bibfnamefont {L.}~\bibnamefont {Li}},
  \bibinfo {author} {\bibfnamefont {L.~E.~F.}\ \bibnamefont {Foa~Torres}},\
  and\ \bibinfo {author} {\bibfnamefont {M.~V.}\ \bibnamefont {Hosseini}},\
  }\bibfield  {title} {\bibinfo {title} {Topological phases of commensurate or
  incommensurate non-hermitian su-schrieffer-heeger lattices},\ }\href
  {https://doi.org/10.1103/PhysRevB.109.205142} {\bibfield  {journal} {\bibinfo
   {journal} {Phys. Rev. B}\ }\textbf {\bibinfo {volume} {109}},\ \bibinfo
  {pages} {205142} (\bibinfo {year} {2024})}\BibitemShut {NoStop}%
\bibitem [{\citenamefont {Yi}\ and\ \citenamefont {Yu}(2001)}]{XXYi_2001}%
  \BibitemOpen
  \bibfield  {author} {\bibinfo {author} {\bibfnamefont {X.~X.}\ \bibnamefont
  {Yi}}\ and\ \bibinfo {author} {\bibfnamefont {S.~X.}\ \bibnamefont {Yu}},\
  }\bibfield  {title} {\bibinfo {title} {Effective hamiltonian approach to the
  master equation},\ }\href {https://doi.org/10.1088/1464-4266/3/6/304}
  {\bibfield  {journal} {\bibinfo  {journal} {J. Opt. B: Quant. Semiclass
  Opt.}\ }\textbf {\bibinfo {volume} {3}},\ \bibinfo {pages} {372} (\bibinfo
  {year} {2001})}\BibitemShut {NoStop}%
\bibitem [{\citenamefont {Naghiloo}\ \emph {et~al.}(2019)\citenamefont
  {Naghiloo}, \citenamefont {Abbasi}, \citenamefont {Joglekar},\ and\
  \citenamefont {Murch}}]{Naghiloo2019}%
  \BibitemOpen
  \bibfield  {author} {\bibinfo {author} {\bibfnamefont {M.}~\bibnamefont
  {Naghiloo}}, \bibinfo {author} {\bibfnamefont {M.}~\bibnamefont {Abbasi}},
  \bibinfo {author} {\bibfnamefont {Y.~N.}\ \bibnamefont {Joglekar}},\ and\
  \bibinfo {author} {\bibfnamefont {K.~W.}\ \bibnamefont {Murch}},\ }\bibfield
  {title} {\bibinfo {title} {Quantum state tomography across the exceptional
  point in a single dissipative qubit},\ }\href
  {https://doi.org/10.1038/s41567-019-0652-z} {\bibfield  {journal} {\bibinfo
  {journal} {Nat. Phys.}\ }\textbf {\bibinfo {volume} {15}},\ \bibinfo {pages}
  {1232} (\bibinfo {year} {2019})}\BibitemShut {NoStop}%
\bibitem [{\citenamefont {Hashemi}\ \emph {et~al.}(2022)\citenamefont
  {Hashemi}, \citenamefont {Busch}, \citenamefont {Christodoulides},
  \citenamefont {Ozdemir},\ and\ \citenamefont {El-Ganainy}}]{Hashemi2022}%
  \BibitemOpen
  \bibfield  {author} {\bibinfo {author} {\bibfnamefont {A.}~\bibnamefont
  {Hashemi}}, \bibinfo {author} {\bibfnamefont {K.}~\bibnamefont {Busch}},
  \bibinfo {author} {\bibfnamefont {D.~N.}\ \bibnamefont {Christodoulides}},
  \bibinfo {author} {\bibfnamefont {S.~K.}\ \bibnamefont {Ozdemir}},\ and\
  \bibinfo {author} {\bibfnamefont {R.}~\bibnamefont {El-Ganainy}},\ }\bibfield
   {title} {\bibinfo {title} {Linear response theory of open systems with
  exceptional points},\ }\href {https://doi.org/10.1038/s41467-022-30715-8}
  {\bibfield  {journal} {\bibinfo  {journal} {Nat. Commun.}\ }\textbf {\bibinfo
  {volume} {13}},\ \bibinfo {pages} {3281} (\bibinfo {year}
  {2022})}\BibitemShut {NoStop}%
\bibitem [{\citenamefont {Lin}\ \emph {et~al.}(2016)\citenamefont {Lin},
  \citenamefont {Pick}, \citenamefont {Lon\ifmmode~\check{c}\else
  \v{c}\fi{}ar},\ and\ \citenamefont {Rodriguez}}]{PhysRevLett.117.107402}%
  \BibitemOpen
  \bibfield  {author} {\bibinfo {author} {\bibfnamefont {Z.}~\bibnamefont
  {Lin}}, \bibinfo {author} {\bibfnamefont {A.}~\bibnamefont {Pick}}, \bibinfo
  {author} {\bibfnamefont {M.}~\bibnamefont {Lon\ifmmode~\check{c}\else
  \v{c}\fi{}ar}},\ and\ \bibinfo {author} {\bibfnamefont {A.~W.}\ \bibnamefont
  {Rodriguez}},\ }\bibfield  {title} {\bibinfo {title} {Enhanced spontaneous
  emission at third-order dirac exceptional points in inverse-designed photonic
  crystals},\ }\href {https://doi.org/10.1103/PhysRevLett.117.107402}
  {\bibfield  {journal} {\bibinfo  {journal} {Phys. Rev. Lett.}\ }\textbf
  {\bibinfo {volume} {117}},\ \bibinfo {pages} {107402} (\bibinfo {year}
  {2016})}\BibitemShut {NoStop}%
\bibitem [{\citenamefont {Hodaei}\ \emph {et~al.}(2017)\citenamefont {Hodaei},
  \citenamefont {Hassan}, \citenamefont {Wittek}, \citenamefont
  {Garcia-Gracia}, \citenamefont {El-Ganainy}, \citenamefont
  {Christodoulides},\ and\ \citenamefont {Khajavikhan}}]{Hodaei2017}%
  \BibitemOpen
  \bibfield  {author} {\bibinfo {author} {\bibfnamefont {H.}~\bibnamefont
  {Hodaei}}, \bibinfo {author} {\bibfnamefont {A.~U.}\ \bibnamefont {Hassan}},
  \bibinfo {author} {\bibfnamefont {S.}~\bibnamefont {Wittek}}, \bibinfo
  {author} {\bibfnamefont {H.}~\bibnamefont {Garcia-Gracia}}, \bibinfo {author}
  {\bibfnamefont {R.}~\bibnamefont {El-Ganainy}}, \bibinfo {author}
  {\bibfnamefont {D.~N.}\ \bibnamefont {Christodoulides}},\ and\ \bibinfo
  {author} {\bibfnamefont {M.}~\bibnamefont {Khajavikhan}},\ }\bibfield
  {title} {\bibinfo {title} {Enhanced sensitivity at higher-order exceptional
  points},\ }\href {https://doi.org/10.1038/nature23280} {\bibfield  {journal}
  {\bibinfo  {journal} {Nature (London)}\ }\textbf {\bibinfo {volume} {548}},\
  \bibinfo {pages} {187} (\bibinfo {year} {2017})}\BibitemShut {NoStop}%
\bibitem [{\citenamefont {Zhou}\ \emph {et~al.}(2023)\citenamefont {Zhou},
  \citenamefont {Ren}, \citenamefont {Xiao}, \citenamefont {Zhang},
  \citenamefont {Huang}, \citenamefont {Li}, \citenamefont {Sun}, \citenamefont
  {Wu}, \citenamefont {Qiu}, \citenamefont {Nori},\ and\ \citenamefont
  {Jing}}]{Zhou2023}%
  \BibitemOpen
  \bibfield  {author} {\bibinfo {author} {\bibfnamefont {X.}~\bibnamefont
  {Zhou}}, \bibinfo {author} {\bibfnamefont {X.}~\bibnamefont {Ren}}, \bibinfo
  {author} {\bibfnamefont {D.}~\bibnamefont {Xiao}}, \bibinfo {author}
  {\bibfnamefont {J.}~\bibnamefont {Zhang}}, \bibinfo {author} {\bibfnamefont
  {R.}~\bibnamefont {Huang}}, \bibinfo {author} {\bibfnamefont
  {Z.}~\bibnamefont {Li}}, \bibinfo {author} {\bibfnamefont {X.}~\bibnamefont
  {Sun}}, \bibinfo {author} {\bibfnamefont {X.}~\bibnamefont {Wu}}, \bibinfo
  {author} {\bibfnamefont {C.-W.}\ \bibnamefont {Qiu}}, \bibinfo {author}
  {\bibfnamefont {F.}~\bibnamefont {Nori}},\ and\ \bibinfo {author}
  {\bibfnamefont {H.}~\bibnamefont {Jing}},\ }\bibfield  {title} {\bibinfo
  {title} {Higher-order singularities in phase-tracked electromechanical
  oscillators},\ }\href {https://doi.org/10.1038/s41467-023-43708-y} {\bibfield
   {journal} {\bibinfo  {journal} {Nat. Commun.}\ }\textbf {\bibinfo {volume}
  {14}},\ \bibinfo {pages} {7944} (\bibinfo {year} {2023})}\BibitemShut
  {NoStop}%
\bibitem [{\citenamefont {Hu}\ \emph {et~al.}(2018)\citenamefont {Hu},
  \citenamefont {Hou}, \citenamefont {Zhang},\ and\ \citenamefont
  {Zhang}}]{PhysRevLett.120.240401}%
  \BibitemOpen
  \bibfield  {author} {\bibinfo {author} {\bibfnamefont {H.}~\bibnamefont
  {Hu}}, \bibinfo {author} {\bibfnamefont {J.}~\bibnamefont {Hou}}, \bibinfo
  {author} {\bibfnamefont {F.}~\bibnamefont {Zhang}},\ and\ \bibinfo {author}
  {\bibfnamefont {C.}~\bibnamefont {Zhang}},\ }\bibfield  {title} {\bibinfo
  {title} {Topological triply degenerate points induced by spin-tensor-momentum
  couplings},\ }\href {https://doi.org/10.1103/PhysRevLett.120.240401}
  {\bibfield  {journal} {\bibinfo  {journal} {Phys. Rev. Lett.}\ }\textbf
  {\bibinfo {volume} {120}},\ \bibinfo {pages} {240401} (\bibinfo {year}
  {2018})}\BibitemShut {NoStop}%
\bibitem [{\citenamefont {Lei}\ \emph {et~al.}(2022)\citenamefont {Lei},
  \citenamefont {Deng},\ and\ \citenamefont {Lee}}]{PhysRevResearch.4.033008}%
  \BibitemOpen
  \bibfield  {author} {\bibinfo {author} {\bibfnamefont {Z.}~\bibnamefont
  {Lei}}, \bibinfo {author} {\bibfnamefont {Y.}~\bibnamefont {Deng}},\ and\
  \bibinfo {author} {\bibfnamefont {C.}~\bibnamefont {Lee}},\ }\bibfield
  {title} {\bibinfo {title} {Unpaired topological triply degenerate point for
  spin-tensor-momentum-coupled ultracold atoms},\ }\href
  {https://doi.org/10.1103/PhysRevResearch.4.033008} {\bibfield  {journal}
  {\bibinfo  {journal} {Phys. Rev. Res.}\ }\textbf {\bibinfo {volume} {4}},\
  \bibinfo {pages} {033008} (\bibinfo {year} {2022})}\BibitemShut {NoStop}%
\bibitem [{\citenamefont {Zhang}\ \emph
  {et~al.}(2022{\natexlab{a}})\citenamefont {Zhang}, \citenamefont {Yuan},
  \citenamefont {Li}, \citenamefont {Luo}, \citenamefont {Liu}, \citenamefont
  {Zhu}, \citenamefont {Qin}, \citenamefont {Zhang}, \citenamefont {Lin},\ and\
  \citenamefont {Du}}]{PhysRevLett.129.250501}%
  \BibitemOpen
  \bibfield  {author} {\bibinfo {author} {\bibfnamefont {M.}~\bibnamefont
  {Zhang}}, \bibinfo {author} {\bibfnamefont {X.}~\bibnamefont {Yuan}},
  \bibinfo {author} {\bibfnamefont {Y.}~\bibnamefont {Li}}, \bibinfo {author}
  {\bibfnamefont {X.-W.}\ \bibnamefont {Luo}}, \bibinfo {author} {\bibfnamefont
  {C.}~\bibnamefont {Liu}}, \bibinfo {author} {\bibfnamefont {M.}~\bibnamefont
  {Zhu}}, \bibinfo {author} {\bibfnamefont {X.}~\bibnamefont {Qin}}, \bibinfo
  {author} {\bibfnamefont {C.}~\bibnamefont {Zhang}}, \bibinfo {author}
  {\bibfnamefont {Y.}~\bibnamefont {Lin}},\ and\ \bibinfo {author}
  {\bibfnamefont {J.}~\bibnamefont {Du}},\ }\bibfield  {title} {\bibinfo
  {title} {Observation of spin-tensor induced topological phase transitions of
  triply degenerate points with a trapped ion},\ }\href
  {https://doi.org/10.1103/PhysRevLett.129.250501} {\bibfield  {journal}
  {\bibinfo  {journal} {Phys. Rev. Lett.}\ }\textbf {\bibinfo {volume} {129}},\
  \bibinfo {pages} {250501} (\bibinfo {year} {2022}{\natexlab{a}})}\BibitemShut
  {NoStop}%
\bibitem [{\citenamefont {Haldane}(1988)}]{PhysRevLett.61.2015}%
  \BibitemOpen
  \bibfield  {author} {\bibinfo {author} {\bibfnamefont {F.~D.~M.}\
  \bibnamefont {Haldane}},\ }\bibfield  {title} {\bibinfo {title} {Model for a
  quantum hall effect without landau levels: Condensed-matter realization of
  the "parity anomaly"},\ }\href {https://doi.org/10.1103/PhysRevLett.61.2015}
  {\bibfield  {journal} {\bibinfo  {journal} {Phys. Rev. Lett.}\ }\textbf
  {\bibinfo {volume} {61}},\ \bibinfo {pages} {2015} (\bibinfo {year}
  {1988})}\BibitemShut {NoStop}%
\bibitem [{\citenamefont {Nagaosa}\ \emph {et~al.}(2010)\citenamefont
  {Nagaosa}, \citenamefont {Sinova}, \citenamefont {Onoda}, \citenamefont
  {MacDonald},\ and\ \citenamefont {Ong}}]{RevModPhys.82.1539}%
  \BibitemOpen
  \bibfield  {author} {\bibinfo {author} {\bibfnamefont {N.}~\bibnamefont
  {Nagaosa}}, \bibinfo {author} {\bibfnamefont {J.}~\bibnamefont {Sinova}},
  \bibinfo {author} {\bibfnamefont {S.}~\bibnamefont {Onoda}}, \bibinfo
  {author} {\bibfnamefont {A.~H.}\ \bibnamefont {MacDonald}},\ and\ \bibinfo
  {author} {\bibfnamefont {N.~P.}\ \bibnamefont {Ong}},\ }\bibfield  {title}
  {\bibinfo {title} {Anomalous hall effect},\ }\href
  {https://doi.org/10.1103/RevModPhys.82.1539} {\bibfield  {journal} {\bibinfo
  {journal} {Rev. Mod. Phys.}\ }\textbf {\bibinfo {volume} {82}},\ \bibinfo
  {pages} {1539} (\bibinfo {year} {2010})}\BibitemShut {NoStop}%
\bibitem [{\citenamefont {Kruthoff}\ \emph {et~al.}(2017)\citenamefont
  {Kruthoff}, \citenamefont {de~Boer}, \citenamefont {van Wezel}, \citenamefont
  {Kane},\ and\ \citenamefont {Slager}}]{PhysRevX.7.041069}%
  \BibitemOpen
  \bibfield  {author} {\bibinfo {author} {\bibfnamefont {J.}~\bibnamefont
  {Kruthoff}}, \bibinfo {author} {\bibfnamefont {J.}~\bibnamefont {de~Boer}},
  \bibinfo {author} {\bibfnamefont {J.}~\bibnamefont {van Wezel}}, \bibinfo
  {author} {\bibfnamefont {C.~L.}\ \bibnamefont {Kane}},\ and\ \bibinfo
  {author} {\bibfnamefont {R.-J.}\ \bibnamefont {Slager}},\ }\bibfield  {title}
  {\bibinfo {title} {Topological classification of crystalline insulators
  through band structure combinatorics},\ }\href
  {https://doi.org/10.1103/PhysRevX.7.041069} {\bibfield  {journal} {\bibinfo
  {journal} {Phys. Rev. X}\ }\textbf {\bibinfo {volume} {7}},\ \bibinfo {pages}
  {041069} (\bibinfo {year} {2017})}\BibitemShut {NoStop}%
\bibitem [{\citenamefont {Bradlyn}\ \emph {et~al.}(2017)\citenamefont
  {Bradlyn}, \citenamefont {Elcoro}, \citenamefont {Cano}, \citenamefont
  {Vergniory}, \citenamefont {Wang}, \citenamefont {Felser}, \citenamefont
  {Aroyo},\ and\ \citenamefont {Bernevig}}]{Bradlyn2017}%
  \BibitemOpen
  \bibfield  {author} {\bibinfo {author} {\bibfnamefont {B.}~\bibnamefont
  {Bradlyn}}, \bibinfo {author} {\bibfnamefont {L.}~\bibnamefont {Elcoro}},
  \bibinfo {author} {\bibfnamefont {J.}~\bibnamefont {Cano}}, \bibinfo {author}
  {\bibfnamefont {M.~G.}\ \bibnamefont {Vergniory}}, \bibinfo {author}
  {\bibfnamefont {Z.}~\bibnamefont {Wang}}, \bibinfo {author} {\bibfnamefont
  {C.}~\bibnamefont {Felser}}, \bibinfo {author} {\bibfnamefont {M.~I.}\
  \bibnamefont {Aroyo}},\ and\ \bibinfo {author} {\bibfnamefont {B.~A.}\
  \bibnamefont {Bernevig}},\ }\bibfield  {title} {\bibinfo {title} {Topological
  quantum chemistry},\ }\href {https://doi.org/10.1038/nature23268} {\bibfield
  {journal} {\bibinfo  {journal} {Nature (London)}\ }\textbf {\bibinfo {volume}
  {547}},\ \bibinfo {pages} {298} (\bibinfo {year} {2017})}\BibitemShut
  {NoStop}%
\bibitem [{\citenamefont {Po}\ \emph {et~al.}(2017)\citenamefont {Po},
  \citenamefont {Vishwanath},\ and\ \citenamefont {Watanabe}}]{Po2017}%
  \BibitemOpen
  \bibfield  {author} {\bibinfo {author} {\bibfnamefont {H.~C.}\ \bibnamefont
  {Po}}, \bibinfo {author} {\bibfnamefont {A.}~\bibnamefont {Vishwanath}},\
  and\ \bibinfo {author} {\bibfnamefont {H.}~\bibnamefont {Watanabe}},\
  }\bibfield  {title} {\bibinfo {title} {Symmetry-based indicators of band
  topology in the 230 space groups},\ }\href
  {https://doi.org/10.1038/s41467-017-00133-2} {\bibfield  {journal} {\bibinfo
  {journal} {Nat. Commun.}\ }\textbf {\bibinfo {volume} {8}},\ \bibinfo {pages}
  {50} (\bibinfo {year} {2017})}\BibitemShut {NoStop}%
\bibitem [{\citenamefont {Bender}\ and\ \citenamefont
  {Boettcher}(1998)}]{PhysRevLett.80.5243}%
  \BibitemOpen
  \bibfield  {author} {\bibinfo {author} {\bibfnamefont {C.~M.}\ \bibnamefont
  {Bender}}\ and\ \bibinfo {author} {\bibfnamefont {S.}~\bibnamefont
  {Boettcher}},\ }\bibfield  {title} {\bibinfo {title} {Real spectra in
  non-hermitian hamiltonians having {P}{T} symmetry},\ }\href
  {https://doi.org/10.1103/PhysRevLett.80.5243} {\bibfield  {journal} {\bibinfo
   {journal} {Phys. Rev. Lett.}\ }\textbf {\bibinfo {volume} {80}},\ \bibinfo
  {pages} {5243} (\bibinfo {year} {1998})}\BibitemShut {NoStop}%
\bibitem [{\citenamefont {Bouhon}\ \emph
  {et~al.}(2020{\natexlab{a}})\citenamefont {Bouhon}, \citenamefont {Wu},
  \citenamefont {Slager}, \citenamefont {Weng}, \citenamefont {Yazyev},\ and\
  \citenamefont {Bzdu{\v{s}}ek}}]{Bouhon2020}%
  \BibitemOpen
  \bibfield  {author} {\bibinfo {author} {\bibfnamefont {A.}~\bibnamefont
  {Bouhon}}, \bibinfo {author} {\bibfnamefont {Q.}~\bibnamefont {Wu}}, \bibinfo
  {author} {\bibfnamefont {R.-J.}\ \bibnamefont {Slager}}, \bibinfo {author}
  {\bibfnamefont {H.}~\bibnamefont {Weng}}, \bibinfo {author} {\bibfnamefont
  {O.~V.}\ \bibnamefont {Yazyev}},\ and\ \bibinfo {author} {\bibfnamefont
  {T.}~\bibnamefont {Bzdu{\v{s}}ek}},\ }\bibfield  {title} {\bibinfo {title}
  {Non-abelian reciprocal braiding of weyl points and its manifestation in
  zrte},\ }\href {https://doi.org/10.1038/s41567-020-0967-9} {\bibfield
  {journal} {\bibinfo  {journal} {Nat. Phys.}\ }\textbf {\bibinfo {volume}
  {16}},\ \bibinfo {pages} {1137} (\bibinfo {year}
  {2020}{\natexlab{a}})}\BibitemShut {NoStop}%
\bibitem [{\citenamefont {Bouhon}\ \emph
  {et~al.}(2020{\natexlab{b}})\citenamefont {Bouhon}, \citenamefont
  {Bzdu\ifmmode~\check{s}\else \v{s}\fi{}ek},\ and\ \citenamefont
  {Slager}}]{PhysRevB.102.115135}%
  \BibitemOpen
  \bibfield  {author} {\bibinfo {author} {\bibfnamefont {A.}~\bibnamefont
  {Bouhon}}, \bibinfo {author} {\bibfnamefont {T.~c.~v.}\ \bibnamefont
  {Bzdu\ifmmode~\check{s}\else \v{s}\fi{}ek}},\ and\ \bibinfo {author}
  {\bibfnamefont {R.-J.}\ \bibnamefont {Slager}},\ }\bibfield  {title}
  {\bibinfo {title} {Geometric approach to fragile topology beyond symmetry
  indicators},\ }\href {https://doi.org/10.1103/PhysRevB.102.115135} {\bibfield
   {journal} {\bibinfo  {journal} {Phys. Rev. B}\ }\textbf {\bibinfo {volume}
  {102}},\ \bibinfo {pages} {115135} (\bibinfo {year}
  {2020}{\natexlab{b}})}\BibitemShut {NoStop}%
\bibitem [{\citenamefont {\"Unal}\ \emph {et~al.}(2020)\citenamefont {\"Unal},
  \citenamefont {Bouhon},\ and\ \citenamefont
  {Slager}}]{PhysRevLett.125.053601}%
  \BibitemOpen
  \bibfield  {author} {\bibinfo {author} {\bibfnamefont {F.~N.}\ \bibnamefont
  {\"Unal}}, \bibinfo {author} {\bibfnamefont {A.}~\bibnamefont {Bouhon}},\
  and\ \bibinfo {author} {\bibfnamefont {R.-J.}\ \bibnamefont {Slager}},\
  }\bibfield  {title} {\bibinfo {title} {Topological euler class as a dynamical
  observable in optical lattices},\ }\href
  {https://doi.org/10.1103/PhysRevLett.125.053601} {\bibfield  {journal}
  {\bibinfo  {journal} {Phys. Rev. Lett.}\ }\textbf {\bibinfo {volume} {125}},\
  \bibinfo {pages} {053601} (\bibinfo {year} {2020})}\BibitemShut {NoStop}%
\bibitem [{\citenamefont {Zhang}\ \emph
  {et~al.}(2022{\natexlab{b}})\citenamefont {Zhang}, \citenamefont {Yang},\
  and\ \citenamefont {Fang}}]{Zhang2022}%
  \BibitemOpen
  \bibfield  {author} {\bibinfo {author} {\bibfnamefont {K.}~\bibnamefont
  {Zhang}}, \bibinfo {author} {\bibfnamefont {Z.}~\bibnamefont {Yang}},\ and\
  \bibinfo {author} {\bibfnamefont {C.}~\bibnamefont {Fang}},\ }\bibfield
  {title} {\bibinfo {title} {Universal non-hermitian skin effect in two and
  higher dimensions},\ }\href {https://doi.org/10.1038/s41467-022-30161-6}
  {\bibfield  {journal} {\bibinfo  {journal} {Nat. Commun.}\ }\textbf {\bibinfo
  {volume} {13}},\ \bibinfo {pages} {2496} (\bibinfo {year}
  {2022}{\natexlab{b}})}\BibitemShut {NoStop}%
\bibitem [{\citenamefont {Zhang}\ \emph
  {et~al.}(2023{\natexlab{b}})\citenamefont {Zhang}, \citenamefont {Fang},\
  and\ \citenamefont {Yang}}]{PhysRevLett.131.036402}%
  \BibitemOpen
  \bibfield  {author} {\bibinfo {author} {\bibfnamefont {K.}~\bibnamefont
  {Zhang}}, \bibinfo {author} {\bibfnamefont {C.}~\bibnamefont {Fang}},\ and\
  \bibinfo {author} {\bibfnamefont {Z.}~\bibnamefont {Yang}},\ }\bibfield
  {title} {\bibinfo {title} {Dynamical degeneracy splitting and directional
  invisibility in non-hermitian systems},\ }\href
  {https://doi.org/10.1103/PhysRevLett.131.036402} {\bibfield  {journal}
  {\bibinfo  {journal} {Phys. Rev. Lett.}\ }\textbf {\bibinfo {volume} {131}},\
  \bibinfo {pages} {036402} (\bibinfo {year} {2023}{\natexlab{b}})}\BibitemShut
  {NoStop}%
\bibitem [{\citenamefont {Wu}\ \emph {et~al.}(2016)\citenamefont {Wu},
  \citenamefont {Zhang}, \citenamefont {Sun}, \citenamefont {Xu}, \citenamefont
  {Wang}, \citenamefont {Ji}, \citenamefont {Deng}, \citenamefont {Chen},
  \citenamefont {Liu},\ and\ \citenamefont
  {Pan}}]{doi:10.1126/science.aaf6689}%
  \BibitemOpen
  \bibfield  {author} {\bibinfo {author} {\bibfnamefont {Z.}~\bibnamefont
  {Wu}}, \bibinfo {author} {\bibfnamefont {L.}~\bibnamefont {Zhang}}, \bibinfo
  {author} {\bibfnamefont {W.}~\bibnamefont {Sun}}, \bibinfo {author}
  {\bibfnamefont {X.-T.}\ \bibnamefont {Xu}}, \bibinfo {author} {\bibfnamefont
  {B.-Z.}\ \bibnamefont {Wang}}, \bibinfo {author} {\bibfnamefont {S.-C.}\
  \bibnamefont {Ji}}, \bibinfo {author} {\bibfnamefont {Y.}~\bibnamefont
  {Deng}}, \bibinfo {author} {\bibfnamefont {S.}~\bibnamefont {Chen}}, \bibinfo
  {author} {\bibfnamefont {X.-J.}\ \bibnamefont {Liu}},\ and\ \bibinfo {author}
  {\bibfnamefont {J.-W.}\ \bibnamefont {Pan}},\ }\bibfield  {title} {\bibinfo
  {title} {Realization of two-dimensional spin-orbit coupling for bose-einstein
  condensates},\ }\href {https://doi.org/10.1126/science.aaf6689} {\bibfield
  {journal} {\bibinfo  {journal} {Science}\ }\textbf {\bibinfo {volume}
  {354}},\ \bibinfo {pages} {83} (\bibinfo {year} {2016})}\BibitemShut
  {NoStop}%
\bibitem [{\citenamefont {Bradlyn}\ \emph {et~al.}(2016)\citenamefont
  {Bradlyn}, \citenamefont {Cano}, \citenamefont {Wang}, \citenamefont
  {Vergniory}, \citenamefont {Felser}, \citenamefont {Cava},\ and\
  \citenamefont {Bernevig}}]{doi:10.1126/science.aaf5037}%
  \BibitemOpen
  \bibfield  {author} {\bibinfo {author} {\bibfnamefont {B.}~\bibnamefont
  {Bradlyn}}, \bibinfo {author} {\bibfnamefont {J.}~\bibnamefont {Cano}},
  \bibinfo {author} {\bibfnamefont {Z.}~\bibnamefont {Wang}}, \bibinfo {author}
  {\bibfnamefont {M.~G.}\ \bibnamefont {Vergniory}}, \bibinfo {author}
  {\bibfnamefont {C.}~\bibnamefont {Felser}}, \bibinfo {author} {\bibfnamefont
  {R.~J.}\ \bibnamefont {Cava}},\ and\ \bibinfo {author} {\bibfnamefont
  {B.~A.}\ \bibnamefont {Bernevig}},\ }\bibfield  {title} {\bibinfo {title}
  {Beyond dirac and weyl fermions: Unconventional quasiparticles in
  conventional crystals},\ }\href {https://doi.org/10.1126/science.aaf5037}
  {\bibfield  {journal} {\bibinfo  {journal} {Science}\ }\textbf {\bibinfo
  {volume} {353}},\ \bibinfo {pages} {aaf5037} (\bibinfo {year}
  {2016})}\BibitemShut {NoStop}%
\bibitem [{\citenamefont {Zhu}\ \emph {et~al.}(2016)\citenamefont {Zhu},
  \citenamefont {Winkler}, \citenamefont {Wu}, \citenamefont {Li},\ and\
  \citenamefont {Soluyanov}}]{PhysRevX.6.031003}%
  \BibitemOpen
  \bibfield  {author} {\bibinfo {author} {\bibfnamefont {Z.}~\bibnamefont
  {Zhu}}, \bibinfo {author} {\bibfnamefont {G.~W.}\ \bibnamefont {Winkler}},
  \bibinfo {author} {\bibfnamefont {Q.}~\bibnamefont {Wu}}, \bibinfo {author}
  {\bibfnamefont {J.}~\bibnamefont {Li}},\ and\ \bibinfo {author}
  {\bibfnamefont {A.~A.}\ \bibnamefont {Soluyanov}},\ }\bibfield  {title}
  {\bibinfo {title} {Triple point topological metals},\ }\href
  {https://doi.org/10.1103/PhysRevX.6.031003} {\bibfield  {journal} {\bibinfo
  {journal} {Phys. Rev. X}\ }\textbf {\bibinfo {volume} {6}},\ \bibinfo {pages}
  {031003} (\bibinfo {year} {2016})}\BibitemShut {NoStop}%
\bibitem [{\citenamefont {Luo}\ \emph {et~al.}(2017)\citenamefont {Luo},
  \citenamefont {Sun},\ and\ \citenamefont {Zhang}}]{PhysRevLett.119.193001}%
  \BibitemOpen
  \bibfield  {author} {\bibinfo {author} {\bibfnamefont {X.-W.}\ \bibnamefont
  {Luo}}, \bibinfo {author} {\bibfnamefont {K.}~\bibnamefont {Sun}},\ and\
  \bibinfo {author} {\bibfnamefont {C.}~\bibnamefont {Zhang}},\ }\bibfield
  {title} {\bibinfo {title} {Spin-tensor--momentum-coupled bose-einstein
  condensates},\ }\href {https://doi.org/10.1103/PhysRevLett.119.193001}
  {\bibfield  {journal} {\bibinfo  {journal} {Phys. Rev. Lett.}\ }\textbf
  {\bibinfo {volume} {119}},\ \bibinfo {pages} {193001} (\bibinfo {year}
  {2017})}\BibitemShut {NoStop}%
\end{thebibliography}
%


\section*{Acknowledgements}
\noindent
{This work was supported by the NSFC (Grants No. 12274473 and No. 12135018).}

\section*{Author contributions}
\noindent
YD conceived the idea and supervised the project.  ZL performed the theoretical and numerical calculations. All authors contributed to the discussion of results and manuscript preparation.

\section*{Competing interests}
\noindent
The authors declare no competing interests.
\\

\section*{Additional information}
\noindent
Supplementary information is available for this paper at...
\\



\clearpage

\begin{widetext}


	\setcounter{equation}{0} \setcounter{figure}{0} \setcounter{table}{0} %
	\renewcommand{\theequation}{S\arabic{equation}} \renewcommand{\thefigure}{S%
		\arabic{figure}} \renewcommand{\bibnumfmt}[1]{[S#1]}
\renewcommand{\thesection}{S\Roman{section}}

\begin{center}
\textbf{\large Supplementary Materials for ``Topological Dynamics and Correspondences in Composite Exceptional Rings''}
\end{center}

\maketitle


	\setcounter{equation}{0} \setcounter{figure}{0} \setcounter{table}{0} %
	\renewcommand{\theequation}{S\arabic{equation}} \renewcommand{\thefigure}{S%
		\arabic{figure}} \renewcommand{\bibnumfmt}[1]{[S#1]}
\renewcommand{\thesection}{S\arabic{section}}


\section{Introduction to TDPs with quadratic SVMC and STMC}\label{appA}

In this section, we introduce triply degenerate points (TDPs) in the context of Hermitian band structure and their associated topological charges briefly. TDPs, akin to direct generation of Weyl points, can be described by: $\pm\sum_{\alpha=x,y,z} k_{\alpha}\hat{F}_{\alpha}$~\cite{doi:10.1126/science.aaf5037,PhysRevX.6.031003} known as spin-vector-momentum coupling (SVMC) for spin-1 matrices $\hat{F}_{\alpha}$ analogous to Pauli matrices for spin-$1/2$ systems. This type of TDP exhibits Chern number $\mathcal{C}_n=\mp2n$. In addition to spin vectors, the describing large spin systems require spin tensor, e.g., the irreducible rank-2 spin-quadrupole tensor $\hat{N}_{\alpha\beta}=(\hat{F}_{\alpha}\hat{F}_{\beta}+\hat{F}_{\beta}\hat{F}_{\alpha})-\delta_{\alpha\beta}\mathbf{F}^2$ for spin-1 system~\cite{PhysRevLett.119.193001}. 

Recent studies~\cite{PhysRevLett.120.240401,PhysRevLett.129.250501} have demonstrated spin-tensor-momentum coupling (STMC) inducing topological transitions and another type of TDP with Chern number $\mathcal{C}_n=\mp n$. Moreover, these two types of topological TDPs just hold two nontrivial bands same as twofold band degenerate points. A recent work~\cite{PhysRevResearch.4.033008} has shown the interplay of quadratic SVMC and STMC can lead to TDP with three nontrivial bands, distinct from the conventional twofold band degenerate points.

Specifically, we employ the Hamiltonian in Eq. (1) of the main text without dissipation, to elucidate the topological properties of TDP:
\begin{eqnarray}\label{TDP}
H_{\rm TDP}(\mathbf{k})=\sum_{\alpha=x,y,z}k_{\alpha}\hat{F}_{\alpha}+u_zk_z\hat{F}_{z}^2+v'_zk^2_z\hat{F}_{z}.
\end{eqnarray}
Except to special parameters ($|u_z|=1$ and $v'_z=0$), this Hamiltonian describes an isolated TDP at the origin of momentum space.
The topological properties of TDPs are characterized by the Chern number defined in Eq. (3) of the main text,  and the corresponding phase diagram is depicted in Fig.~\ref{supfig_TDP}. When $0<u_z<1$, TDP exhibits Chern number $\mathcal{C}_n=-2n$, akin to those arising from SVMC. Conversely, TDP hosts Chern number $\mathcal{C}_n=-n$ as $|u_z|>1$. When $u_z=1$ and $v'_z\neq0$, all three bands of TDP are nontrivial, characterized by the set of Chern numbers $\{2,-1,-1\}$. Notably, the lower (upper) two bands are degenerate along the positive (negative) $z$ axis when $|u_z|=1$ and $v'_z=0$, leading to an ill-defined Chern number in described in Eq. (3) of the main text.

\begin{figure}[!htp]
\includegraphics[width=0.9\columnwidth]{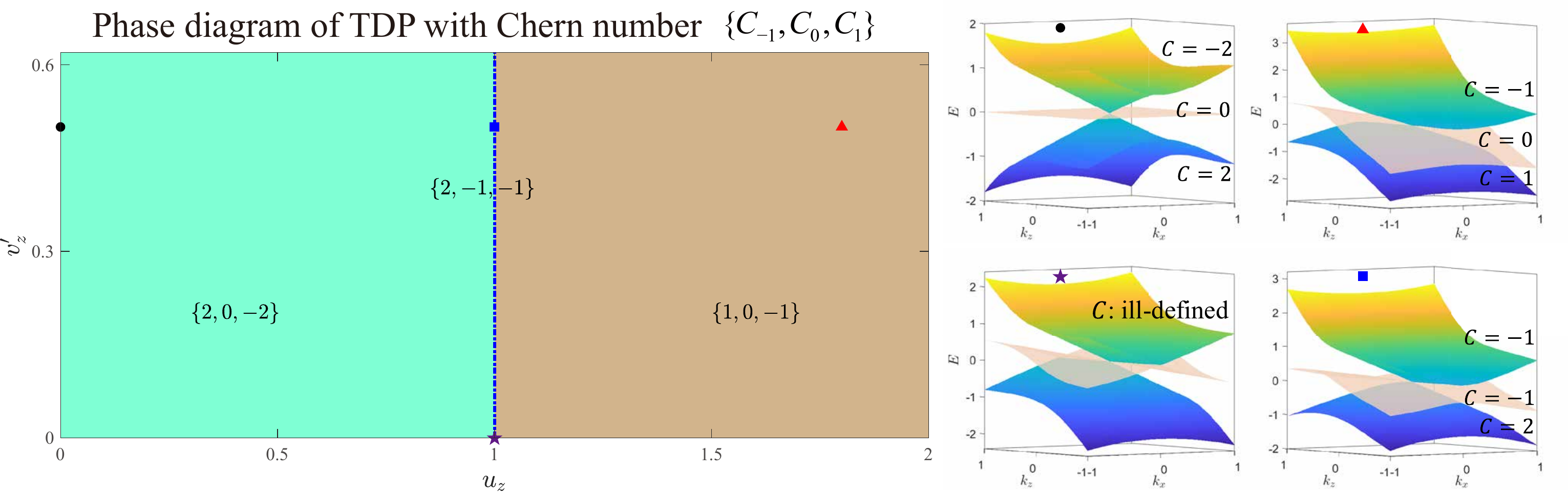}
\caption{\label{supfig_TDP} \textbf{The phase diagram with Chern number of TDP and typical energy bands in $k_y=0$ plane} The set of Chern number $\{\mathcal{C}_{-1},\mathcal{C}_{0},\mathcal{C}_{1}\}$ is provided for distinct phases. The energy bands in the $k_y=0$ plane are shown to the right of the phase diagram for parameters $(u_z=0,v'_z=0.5)$, $(u_z=1.8,v'_z=0.5)$, $(u_z=1,v'_z=0)$, and $(u_z=1,v'_z=0.5)$. Here, two of three bands are degenerate along the $z$ axis when $u_z=1$ and $v'_z=0$, rendering the associated Chern number ill-defined.}
\end{figure}

The phase diagram of TDP seems similar to the one for composite exceptional ring given in Fig. 1(d) of the main text. However, the Chern number in Fig.~\ref{supfig_TDP} describes the topological properties of single TDP, while the change of Chern number in Fig.1(d) of the main text is primarily result from the splitting and vanishing of Weyl exceptional rings (WERs).

\section{The evolution of composite exceptional ring}\label{appB}

In this section, we present additional numerical results to validate the evolution of composite exceptional ring (CER) illustrated in Figs. 1(b) and 1(c) of the main text.

We explore the relationship between symmetries and the formation of exceptional structure. The appearance of exceptional structure can be extracted by resultants derived from the characteristic polynomial of the Hamiltonian. The polynomial $p(\lambda)=\det[\lambda\hat{F}_0-H(\mathbf{k})]$ is expressed as
\begin{eqnarray}\label{POL}
\lambda^3+a_2\lambda^2+a_1\lambda+a_0,
\end{eqnarray}
where the coefficients are given by  $a_2=-2u_zk_z$, $a_1=u_z^2k^2_z-(1+v'_zk_z)^2k^2_z+\gamma^2-k_x^2-k_y^2-2i\gamma(1+v'_zk_z)k_z$, and $a_0=u_zk_z(k_x^2+k_y^2)$. Obviously, this polynomial is invariants under transition $(k_x,k_y)\rightarrow(\cos\phi k_x-\sin\phi k_y,\sin\phi k_x+\cos\phi k_y)$ according with the rotation symmetry.

The polynomial $p(\lambda)$ as well as its first derivative [$p'(\lambda)$] and second derivative [$p''(\lambda)$] vanish at 3EP, while only $p(\lambda)$ and $p'(\lambda)$ vanish at 2EPs. The conditions for these EPs can be well solved by using resultants $R_{p,p'}\equiv\det(S_{p,p'})$ and $R_{p,p''}\equiv\det(S_{p,p''})$, where $S_{P_a,P_b}$ denotes the Sylvester matrix of polynomial $P_a$ and $P_b$. Resultants $R_{P_a,P_b}$ will vanish when $P_a$ and $P_b$ share a common root. Therefor the appearance of 3EP with eigenvalue $\lambda_0$ associating with $R_{p,p'}(\lambda_0)=0$ and $R_{p,p''}(\lambda_0)=0$ while just $R_{p,p'}(\lambda_0)=0$ occurs for 2EP~\cite{PhysRevLett.127.186602}.

In general, these resultants are complex functions of $\lambda$, and thus the codimension of a 3EP (2EP) is typically $4$ ($2$). However, the presence of $\mathcal{PT}$ symmetry and pseudochirality symmetry~\cite{Wu2024} can reduce the codimension of 3EP to $1$. Specifically, $\mathcal{PT}$ symmetry ensures that the coefficients of $p(\lambda)$ in Eq.~\ref{POL} as well as resultants are real functions, reducing the codimension of 3EP to $2$ under constrain of this symmetry. Furthermore, under pseudochirality symmetry, coefficient $a_i$ with odd (even)  indices are purely real (imaginary). Consequently, conditions like $a_0=a_2=0$ indicate that $R_{p,p''}=0$ spontaneously under these two symmetries, further reducing the codimension of 3EP to $1$.

In the Hamiltonian $H$ in Eq.(1) of the main text, these symmetries manifest in the $k_z=0$ plane, where a 1D structure composed of 3EPs can be induced in this 2D space. Indeed, in this plane $p(\lambda)=\lambda^3+(\gamma^2-k_x^2-k_y^2)\lambda=0$, indicating the appearance of 3EPs with a coalescing eigenvalue $\lambda_0=0$ when $k_x^2+k_y^2=\gamma^2$, as discussed in the main text. Outside of this plane,  $\mathcal{PT}$ symmetry and pseudochirality symmetry are no long applicable, and 3EP does not appear, restoring their codimension recovers to $4$. However, a 1D structure made up of 2EPs may appear in this 3D space with a codimension of $2$ for 2EP. The condition $R_{p,p'}=0$ gives rise to
\begin{eqnarray}\label{2EPRe1}
(27a_0+4a_2^3)a_0+(-18a_0a_2+4a_1^2-a_1a_2^2)a_1=0.
\end{eqnarray}
Obviously,  for $u_z=0$ ($a_2=a_0=0$),  the above equation is satisfied only when $a_1=0$, confirming that 2EP actually manifest as 3EP. Thereby, the 2EP does not appear when STMC vanishes ($u_z=0$), as discussed in the main text.
Furthermorte, for $u_z\neq0$ and $v'_z=0$ (where quadratic SVMC vanishes), the transformations $a_2\rightarrow-a_2$, $a_1\rightarrow a^*_1$ and $a_0\rightarrow-a_0$ under transition $k_z\rightarrow-k_z$ ensure that the left side of Eq.~\eqref{2EPRe1} just becomes its complex conjugate, so the 2EPs (if exist) will be symmetric about $k_z=0$ plane in this situation confirmed the symmetry analysis in main text. The appearance and location of 2EPs can be obtained numerically by solving Eq.~\eqref{2EPRe1}, or alternatively by calculating the energy gap as detailed in the following discussions.

\begin{figure}[!htp]
\includegraphics[width=0.9\columnwidth]{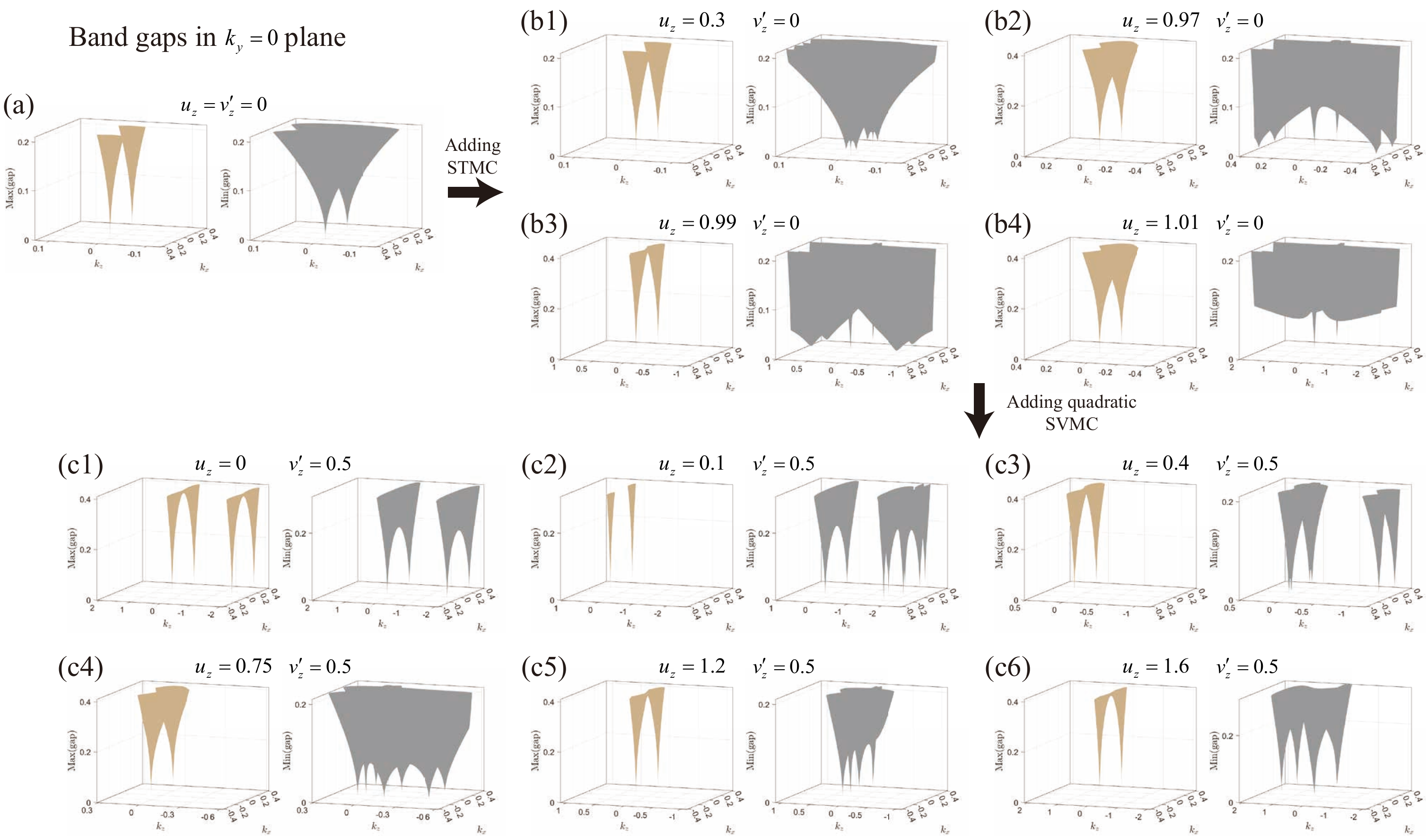}
\caption{\label{supfig_checkgap} \textbf{Verification of exceptional rings locations through gap functions in $k_y=0$ plane.} The quantities $\rm Max(\rm gap)$ and $\rm Min(\rm gap)$ defined in Eq.\eqref{gap} are depicted. The positions of 2EPs (3EPs) can be determined by $\rm Max(\rm gap)\neq0$ and $\rm Min(\rm gap)=0$ ($\rm Max(\rm gap)=\rm Min(\rm gap)=0$). Each panel displays $\gamma=0.2$ with additional parameters specified at the top of each panel.}
\end{figure}

Furthermore, we verify the presence and location of exceptional rings by analyzing the energy gap characteristics of the Hamiltonian $H$ in Eq. (1) in the main text. Specifically focusing on the $k_y=0$ plane, the energy gap quantities are defined as
\begin{eqnarray}\label{gap}
	\rm Max(\rm gap)&=&\max(|E_1-E_0|,|E_0-E_{-1}|,|E_1-E_{-1}|),\nonumber\\
\rm Min(\rm gap)&=&\min(|E_1-E_0|,|E_0-E_{-1}|,|E_1-E_{-1}|),
\end{eqnarray}
where $E_i$ represents the eigenenergies of $H$ at each momentum point $\bf k$. Obviously, $\rm Max(\rm gap)=\rm Min(\rm gap)=0$ associate with third-order exceptional point (3EP), while $\rm Max(\rm gap)\neq0$ and $\rm Min(\rm gap)=0$ signify a second-order exceptional point (2EP).
Utilizing rotation symmetry, two symmetric 3EPs and 2EPs in the $k_y=0$ plane correspond to a third-order exceptional ring (TER) and WER, respectively, in 3D space.

For the simplest scenario with vanishing STMC and quadratic SVMC $u_z=v'_z=0$, $\rm Min(\rm gap)=\rm Max(\rm gap)=0$ occurs at two momenta $(\pm\gamma,0,0)$ as shown in Fig.~\ref{supfig_checkgap}(a).  This confirms the presence of a TER due to the rotational symmetry $\hat{\mathcal{C}}_{z}$. Upon introducing STMC, $\rm Min(\rm gap)=\rm Max(\rm gap)=0$ can occurs at $(\pm\gamma,0,0)$, while $\rm Min(\rm gap)=0$ and $\rm Max(\rm gap)\neq0$ can occur at another four momenta with a symmetric structure $(\pm k_x,0,\pm k_z)$ in addition, as shown in Fig.~\ref{supfig_checkgap}(b1). This result indicates the TER separates into one TER and two WERs upon the introduction of STMC. Furthermore, as the strength of STMC ($u_z$) increases, WERs will move away from TER (specifically, $|k_z|$ of WERs increases), as depicted in Figs.~\ref{supfig_checkgap}(b1-b3). When $u_z>1$, this pair of WERs vanishes immediately, as shown in Fig.~\ref{supfig_checkgap}(b4), indicating that WERs exist within the range $k_z\in(-1,1)$. This sensitivity of WERs location to STMC strength suggests potential applications in precision measurement.

The evolution of exceptional rings becomes more complex within the inclusion of quadratic SVMC ($u_z\neq0$ and $v'_z\neq0$). Two effects are observed after introducing quadratic SVMC. Firstly, another TER appears with non-vanishing quadratic SVMC and vanishing STMC ($v'_z\neq0$ and $u_z=0$), as shown in Fig.~\ref{supfig_checkgap}(c1). Unlike the TER in the $k_z=0$ plane, this TER vanishes when the STMC is added.
Specifically, it will separate into three WERs when $v'_z\neq0$ and $u_z\neq0$, as shown in Fig.~\ref{supfig_checkgap}(c2). Moreover, one of these WERs will get closer to TER (specifically $|k_z|$ becomes smaller) with increasing $u_z$, as displayed in Fig.~\ref{supfig_checkgap}(c3).
Secondly, these two WERs separated from the TER in $k_z=0$ plane will not be in response to STMC equally because of the broken of $\mathcal{CM}_{x}$ by quadratic SVMC. In Fig.~\ref{supfig_checkgap}(c4), these two WERs will be asymmetric about $k_z=0$ plane obviously due to the finite strength of quadratic SVMC.

Combining these effects, two of these WERs, which are originally from the TER in $k_z=0$ and $k_z\neq0$ plane, respectively, will become closer, as demonstrated in Figs.~\ref{supfig_checkgap}(c3) and \ref{supfig_checkgap}(c4). After these two WERs merging, they eventually annihilate with each other indicating they hold opposite topological charges, as shown in Fig.~\ref{supfig_checkgap}(c5). In this case, there is just one WER remains in the range WER living for without quadratic SVMC, and it will also leave, when $u_z>1+v'_z$, as shown in Fig.~\ref{supfig_checkgap}(c6). In summary, the intricate evolution of exceptional rings discussed here aligns with the complexities illustrated in Fig.1(c) of the main text.

\begin{figure}[!htp]
\includegraphics[width=0.8\columnwidth]{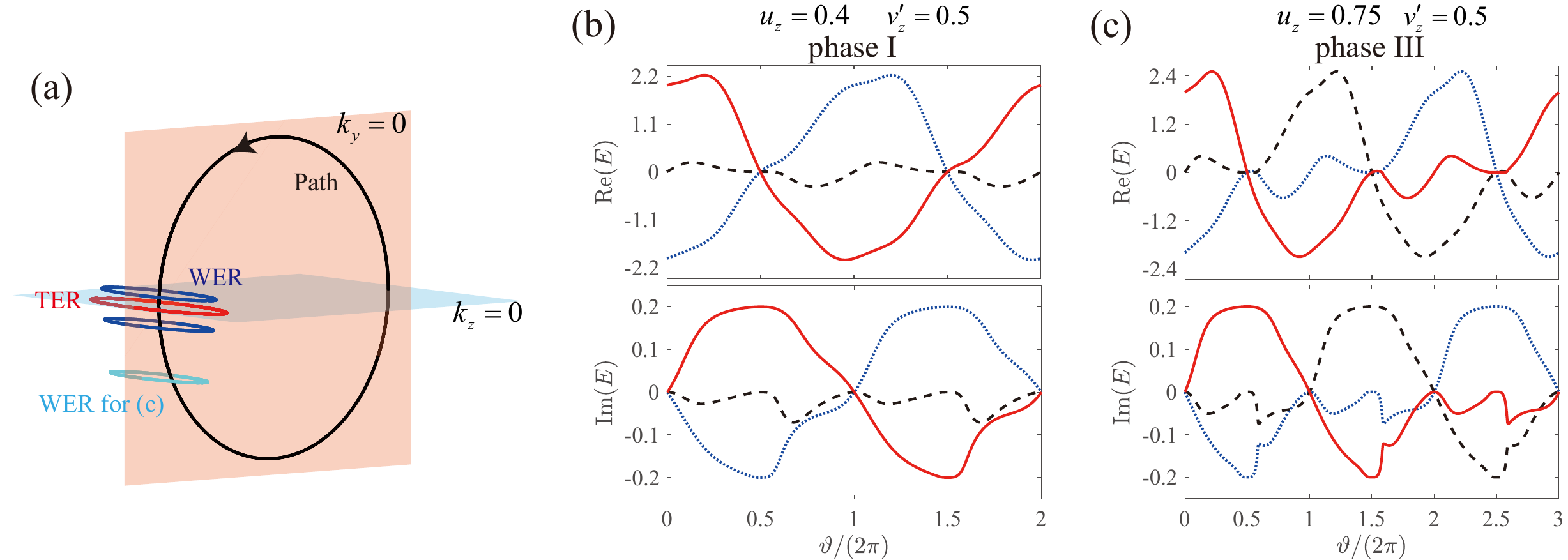}
\caption{\label{supfig_circle_moreWERs} \textbf{The instantaneous spectra along fixed path when circled CER contains more WERs.} (a) The fixed closed path defined in Eq.5 of main text and the location of exceptional rings. Comparing to Fig.2 of the main text, this path will circle additional pair of WER. (b-c) The spectra along this path for different situations, associating with phase {\rm I} and {\rm III}, respectively. The parameter is $v'_{z}=0.5$ and $\gamma=0.2$ in all panels, while $u_{z}=0.4$ for (b), $0.75$ for (c).}
\end{figure}

\section{More examples of encircling processes}\label{appD}

In this section, we present additional results from encircling processes using parameters different from those discussed in Fig. 2 and Fig. 3 of the main text. These parameters lead to variations in the components of CER, resulting in the presence of more WERs, as illustrated in Fig.~\ref{supfig_circle_moreWERs}(a).

Despite the variation in the number of WERs, the Chern number remains the same as in Figs. 2(b) and 2(d) of the main text. Consequently, the quasistatic spectrum also exhibits the same periodicity, as demonstrated in Figs.~\ref{supfig_circle_moreWERs}(b) and ~\ref{supfig_circle_moreWERs}(c). Specifically, the quasistatic spectrum displays double period and the unbraided band with a middle imaginary energy as shown in Fig.~\ref{supfig_circle_moreWERs}(b), which associates with the Chern number $\mathcal{C}=-2n$ of CER. For a Chern number configuration of $\{2,-1,-1\}$,  although the CER contains three WERs, the quasistatic spectrum displays a triple period, as depicted in Fig.~\ref{supfig_circle_moreWERs}(c).

Furthermore, the corresponding encircling dynamic results under these parameters align with those in Figs. 3(b) and 3(d) of the main text. For a CER with Chern number $\mathcal{C}=-2n$ and $\gamma>0$, the state evolves into the branch with the lowest and highest real energies after one period in counterclockwise (CCW) and clockwise (CW) dynamics, respectively, i.e., $|\varphi^{(+)}_{\circlearrowleft}(T)\rangle=|\psi_{-1}(\mathbf{k}_{0})\rangle$ and  $|\varphi^{(+)}_{\circlearrowright}(T)\rangle=|\psi_{1}(\mathbf{k}_{0})\rangle$ shown in Fig.~\ref{supfig_dynamic_moreWERs}(a). Conversely, when the sign of dissipation is reversed, the results for CCW and CW dynamics are swapped: $|\varphi^{(-)}_{\circlearrowleft}(T)\rangle=|\psi_{1}(\mathbf{k}_{0})\rangle$ and  $|\varphi^{(-)}_{\circlearrowright}(T)\rangle=|\psi_{-1}(\mathbf{k}_{0})\rangle$. With a Chern number configuration of $\{2,-1,-1\}$, the CCW (CW) result is $|\varphi^{(+)}_{\circlearrowleft}(T)\rangle=|\psi_{-1}(\mathbf{k}_{0})\rangle$   ($|\varphi^{(+)}_{\circlearrowright}(T)\rangle=|\psi_{1}(\mathbf{k}_{0})\rangle$) for $\gamma>0$ and $|\varphi^{(-)}_{\circlearrowleft}(T)\rangle=|\psi_{0}(\mathbf{k}_{0})\rangle$   ($|\varphi^{(-)}_{\circlearrowright}(T)\rangle=|\psi_{-1}(\mathbf{k}_{0})\rangle$) for $\gamma<0$, as shown in Fig.~\ref{supfig_dynamic_moreWERs}(b).
Therefore, the results of encircling dynamic are also determined by the Chern numbers of CER rather than by the specific components comprising it.

\begin{figure}[!htp]
\includegraphics[width=0.8\columnwidth]{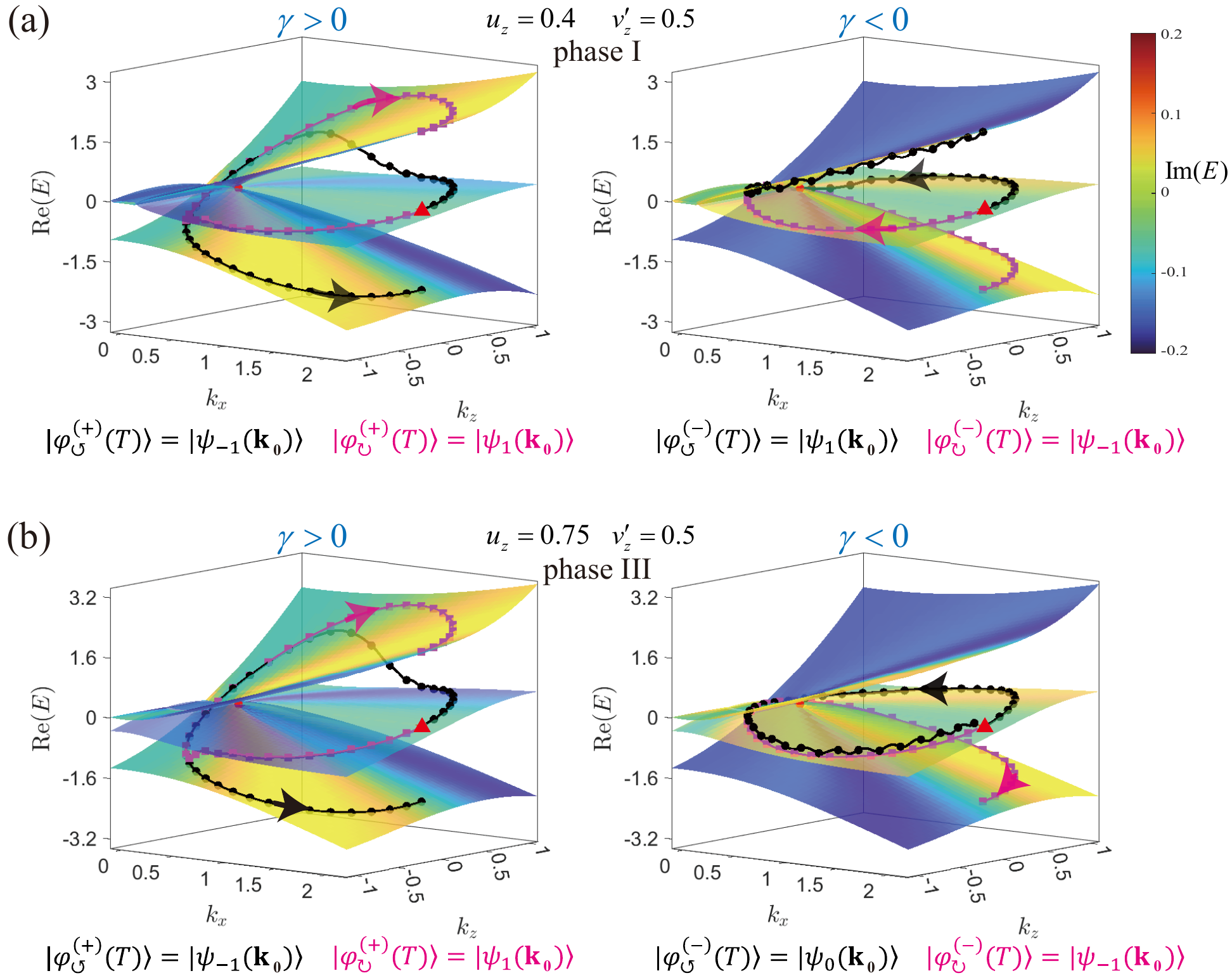}
\caption{\label{supfig_dynamic_moreWERs} \textbf{Encirclement dynamics for circled CER containing different WERs.} In (a), CER contains a TER and pair of WERs with Chern number $\mathcal{C}=-2n$. (b) shows an additional WER circled, resulting in Chern number $\{2,-1,-1\}$. The expected energy in the dynamical process with CCW (CW) direction is plotted by black line with circles (magenta line with squares). The energy of initial state is indexed by a red triangle. Parameters are chosen as $v'_{z}=0.5$, $|\gamma|=0.2$ and $|\omega|=0.02\pi$ for all panels, within $u_{z}=0.4$ for (a) and $0.75$ for (b).}
\end{figure}

\end{widetext}
\end{document}